%% file: main.tex
\newif\ifarxiv
\arxivtrue

\documentclass[journal,12pt,letterpaper,twopage]{IEEEtran}
\IEEEoverridecommandlockouts

\usepackage[displaymath]{lineno}
\usepackage{cite}
\usepackage[cmex10]{amsmath}
\usepackage{amssymb,amsfonts}
\usepackage{algorithmic}
\usepackage{graphicx}
\usepackage{textcomp}
\usepackage{xcolor}
\def\BibTeX{{\rm B\kern-.05em{\sc i\kern-.025em b}\kern-.08em
    T\kern-.1667em\lower.7ex\hbox{E}\kern-.125em}}

\usepackage[utf8]{inputenc}
\usepackage{mleftright}       %
\mleftright                   %

\usepackage{amsthm}
\usepackage{listings}
\usepackage{pgfplots}

\usetikzlibrary{arrows.meta,backgrounds,positioning,patterns}
\usepgfplotslibrary{patchplots,fillbetween}
\pgfplotsset{%
layers/standard/.define layer set={%
    background,axis background,axis grid,axis ticks,axis lines,axis tick labels,pre main,main,axis descriptions,axis foreground%
}{grid style= {/pgfplots/on layer=axis grid},%
    tick style= {/pgfplots/on layer=axis ticks},%
    axis line style= {/pgfplots/on layer=axis lines},%
    label style= {/pgfplots/on layer=axis descriptions},%
    legend style= {/pgfplots/on layer=axis descriptions},%
    title style= {/pgfplots/on layer=axis descriptions},%
    colorbar style= {/pgfplots/on layer=axis descriptions},%
    ticklabel style= {/pgfplots/on layer=axis tick labels},%
    axis background@ style={/pgfplots/on layer=axis background},%
    3d box foreground style={/pgfplots/on layer=axis foreground},%
    },
}
\pgfplotsset{compat=1.18}
\pgfdeclarelayer{standard}
\pgfsetlayers{standard,main}

\usepackage{xspace}
\usepackage[ruled,vlined]{algorithm2e}
\usepackage{tabularx}
\usepackage{booktabs}
\usepackage{arydshln}
\usepackage[normalem]{ulem}
\usepackage{placeins}
\usepackage{balance}
\usepackage[capitalize]{cleveref}
\usepackage{siunitx}
\usepackage{intcalc}
\usepackage[inline]{enumitem}
\usepackage{float}

\makeatletter
\newcommand{\removelatexerror}{\let\@latex@error\@gobble}
\makeatother

\definecolor{DarkGreen}{rgb}{0.1,0.5,0.1}
\definecolor{DarkRed}{rgb}{0.5,0.1,0.1}
\definecolor{DarkBlue}{rgb}{0.1,0.1,0.5}

\newtheorem{theorem}{Theorem}
\newtheorem{corollary}{Corollary}

\newtheorem{definition}{Definition}

\newtheorem{remark}{Remark}

\input{commands}

\begin{document}
\ifarxiv
\else
    \linenumbers
\fi

\title{Byzantine-Resilient Gradient Coding through Local Gradient Computations}

\author{
    \IEEEauthorblockN{Christoph Hofmeister, Luis Maßny, Eitan Yaakobi, and Rawad Bitar}
    \thanks{This paper was presented in part at the 2023 IEEE International Symposium on Information Theory (ISIT)~\cite{hofmeister2023TradingCommunication}.}
    \thanks{CH, LM, and RB are with the School of Computation, Information and Technology at the Technical University of Munich, Germany. Emails: \{christoph.hofmeister, luis.massny, rawad.bitar\}@tum.de}
    \thanks{EY is with the CS department of Technion---Israel Institute of Technology, Israel. Email: yaakobi@cs.technion.ac.il}
    \thanks{This project is funded by the Bavarian Ministry of Economic Affairs, Regional Development and Energy within the scope of the 6G Future Lab Bavaria, and by DFG (German Research Foundation) projects under Grant Agreement No. WA 3907/7-1 and No. BI 2492/1-1.
    This project has received funding from the European Research Council (ERC) under the European Union’s Horizon 2020 research and innovation programme (grant agreement No. 801434).}
}

\ifarxiv
    \markboth{Submitted to the IEEE Trans. Inform. Theory., January 2024}
\fi

\maketitle

\begin{abstract}
  We consider gradient coding in the presence of an adversary controlling so-called malicious workers trying to corrupt the computations. Previous works propose the use of MDS codes to treat the responses from malicious workers as errors and correct them using the error-correction properties of the code. This comes at the expense of increasing the replication, i.e., the number of workers \emph{each partial gradient} is computed by. In this work, we propose a way to reduce the replication to $\nmalicious+1$ instead of $2\nmalicious+1$ in the presence of $\nmalicious$ malicious workers. Our method detects erroneous inputs from the malicious workers, transforming them into erasures. This comes at the expense of $\nmalicious$ additional local computations at the \master and additional rounds of light communication between the \master and the workers. We define a general framework and give fundamental limits for fractional repetition data allocations. Our scheme is optimal in terms of replication and local computation and incurs a communication cost that is asymptotically, in the size of the dataset, a multiplicative factor away from the derived bound. We furthermore show how additional redundancy can be exploited to reduce the number of local computations and communication cost, or, alternatively, tolerate straggling workers.
\end{abstract}


\ifarxiv \else \pagebreak \fi
\input{Journal/sections/introduction}
\input{Journal/sections/problem_setting}

\input{Journal/sections/toy_example_extended}
\input{Journal/sections/main_results}

\input{Journal/sections/framework_general}
\input{Journal/sections/fundamentals}

\input{Journal/sections/scheme}

\input{Journal/sections/conclusion}

\appendix

\subsection{Symmetrization Attack Table}
\label{app:large_table}
\input{Journal/sections/large_table}

\subsection{Algorithms}
\label{app:algorithms}
\input{Journal/sections/algorithms}

\bibliographystyle{IEEEtran}
\bibliography{references}

\end{document}

%% file: commands.tex
\DeclareMathOperator*{\argmin}{arg\,min}
\newcommand{\lis}[1]{\ensuremath{\mathfrak{#1}}}
\newcommand{\setstyle}[1]{\ensuremath{\mathcal{#1}}}
\newcommand{\compbound}{\ensuremath{\left\lfloor \frac{\nmalicious}{\nhonest}\right\rfloor}\xspace}
\newcommand{\nworker}{\ensuremath{n}\xspace}
\newcommand{\ngroup}{\ensuremath{m}\xspace}
\newcommand{\ngrad}{\ensuremath{p}\xspace}
\newcommand{\graddim}{\ensuremath{d}\xspace}
\newcommand{\tgrad}{\ensuremath{\mathbf{g}}\xspace}
\newcommand{\gestim}{\ensuremath{\mathbf{\widehat{g}}}\xspace}
\newcommand{\pgrad}[1][\gind]{\ensuremath{\mathbf{g}_{#1}}\xspace}
\newcommand{\wpgrad}[2]{\ensuremath{{\mathbf{\widetilde{g}}^{(#2)}_{#1}}}\xspace}
\newcommand{\exwpgrad}[2]{\ensuremath{\widetilde{g}^{(#2)}_{#1}}\xspace}

\newcommand{\gind}{\ensuremath{i}\xspace}
\newcommand{\wind}{\ensuremath{j}\xspace}
\newcommand{\compind}{\ensuremath{\zeta}\xspace}
\newcommand{\sample}[1][\gind]{\ensuremath{\mathbf{x}_{#1}}\xspace}
\newcommand{\worker}[1][\wind]{\ensuremath{W_{#1}}\xspace}
\newcommand{\galpha}{\ensuremath{\setstyle{A}}\xspace}

\newcommand{\nmalicious}{\ensuremath{s}\xspace}
\newcommand{\encfun}[1][]{\ensuremath{\operatorname{enc}_{#1}}\xspace}
\newcommand{\encind}{\ensuremath{e}\xspace}
\newcommand{\encdim}[2]{{d_{#1,#2}}\xspace}
\newcommand{\encfunset}{\ensuremath{\lis{E}}\xspace}
\newcommand{\decfun}[1][]{\ensuremath{\operatorname{dec}_{#1}}\xspace}
\newcommand{\nencfun}{\ensuremath{v}\xspace}
\newcommand{\replfact}[1][]{\ensuremath{\rho_{\textrm{#1}}}\xspace}
\newcommand{\commoh}[1][]{\ensuremath{\kappa_{\textrm{#1}}}\xspace}
\newcommand{\proto}{\ensuremath{\Pi}\xspace}
\newcommand{\nround}[1][]{\ensuremath{r_{#1}}\xspace}
\newcommand{\localcomp}[1][]{\ensuremath{c_{#1}}\xspace}
\newcommand{\elimset}{\ensuremath{{\setstyle{S}}}\xspace}
\newcommand{\params}{\ensuremath{\boldsymbol{\theta}}\xspace}
\newcommand{\learningrate}{\ensuremath{\eta^{(\gditer)}}\xspace}
\newcommand{\loss}{\ensuremath{\ell}\xspace}
\newcommand{\gditer}{\ensuremath{\tau}\xspace}
\newcommand{\range}[1]{\ensuremath{[#1]}\xspace}
\newcommand{\roundind}{\ensuremath{t}\xspace}
\newcommand{\respdim }[2]{\ensuremath{d_{{#2},{#1}}}\xspace}
\newcommand{\iresp}{\ensuremath{\mathbf{z}}\xspace}
\newcommand{\noisyiresp}{\ensuremath{\mathbf{\widetilde{z}}}\xspace}

\newcommand{\irespset}{\ensuremath{\lis{Z}}\xspace}
\newcommand{\recresset}{\ensuremath{\widetilde{\irespset}}\xspace}
\newcommand{\workerset}{\ensuremath{\setstyle{W}}\xspace}
\newcommand{\locgradset}{\ensuremath{\lis{G}}\xspace}
\newcommand{\lgindset}{\ensuremath{\lis{I}}\xspace}
\newcommand{\disagreegradset}{\ensuremath{\setstyle{\widetilde{I}}}\xspace}
\newcommand{\allocmat}[1][]{\ensuremath{\mathbf{A}_{#1}}}
\newcommand{\wtmdata}{\ensuremath{\setstyle{D}}\xspace}
\newcommand{\I}{\ensuremath{\mathrm{I}}\xspace}
\newcommand{\En}{\ensuremath{\mathrm{H}}\xspace}
\newcommand{\given}{\;\!\vert\;\!}
\newcommand{\card}[1]{{\left\vert #1 \right\vert}}

\newcommand{\defeq}{\overset{\text{\tiny def}}{=}}
\newcommand{\master}{main node\xspace}
\newcommand{\BGC}{$\nmalicious$-BGC\xspace}
\newcommand{\R}{\mathbb{R}}
\newcommand{\N}{\mathbb{N}}
\newcommand{\playA}{Alice\xspace}
\newcommand{\playB}{Bob\xspace}
\newcommand{\playC}{Carol\xspace}
\newcommand{\playD}{Frank\xspace}
\newcommand{\playboss}{Dan\xspace}
\newcommand{\playbossletter}{D\xspace}
\newcommand{\playbosspronoun}{he\xspace}
\newcommand{\playbossPronoun}{He\xspace}
\newcommand{\nhonest}{\ensuremath{u}}
\newcommand{\nstraggle}{\ensuremath{{\nhonest^\prime}}}
\newcommand{\indone}{{\wind_1}}
\newcommand{\indtwo}{{\wind_2}}
\newcommand{\indcheck}{{\gind_\mathrm{check}}}
\newcommand{\groupset}{\ensuremath{\mathcal{G}}}
\newcommand{\voteset}{\ensuremath{\mathcal{V}}}
\newcommand{\comp}[2][\compind]{{\left[ {#2} \right]_{#1}}}

%% file: Journal/sections/introduction.tex
\section{Introduction}
\label{sec:introduction}

\IEEEPARstart{C}{onsider} the training of a machine learning model using gradient descent at a central processor with access to a large dataset. We refer to this central processor as the main node. In order to speed up the training, it is a common practice to distribute the computation load to a set of worker nodes~\cite{abadi2016tensorflow}, or workers for short.
Among the main vulnerabilities of distributed gradient descent are faulty computations and adversarial attacks: even a single corrupted computation result can drastically deteriorate the performance of the algorithm ~\cite{damaskinos2019aggregathor,blanchardMachineLearningAdversaries}. Consider the popular model of Byzantine errors causing unpredictable malicious computation results~\cite{lamportByzantineGeneralsProblem}. We refer to workers that encounter Byzantine errors as malicious workers and consider a setting in which an adversary can fully control up to a certain number of malicious workers. In this setting, the goal is to develop Byzantine-resilient techniques for distributed gradient descent.

The problem of Byzantine-resilient distributed computations has been studied in a variety of settings. For the case of linear computations, such as matrix-matrix and matrix-vector multiplications, the use of Freivalds' algorithm was proposed to detect malicious responses with high probability~\cite{hofmeisterSecurePrivateAdaptive2022,tangAdaptiveVerifiableCoded2022}. Furthermore, for linear computations, it is possible to encode the computation inputs by linear error correcting codes, exploiting redundancy among the workers' computations~\cite{dataDataEncodingByzantineResilient2021}. Other works have applied methods from group testing~\cite{solankiNonColludingAttacksIdentification2019} and homomorphic hash functions~\cite{keshtkarjahromiSecureCodedCooperative2019} to the linear matrix-vector multiplication problem. For matrix-matrix multiplications, polynomial codes have emerged as a popular choice~\cite{subramaniam2019}. For the larger class of polynomial functions, a coding-theoretic solution is to encode the computation input data by polynomial evaluations~\cite{yuLagrangeCodedComputing2019}.

In the field of distributed machine learning, the functions that need to be computed are usually gradients of highly non-linear functions; that is, methods developed for linear and polynomial function computation are not applicable. There have been many efforts to tackle the problem of Byzantine workers using a variety of methods.
Contributions from both the field of optimization theory and coding theory provided Byzantine-resilient distributed gradient descent methods. For instance, the works in~\cite{yinByzantineRobust2018,chenDistributedStatisticalMachine2017,alistarhByzantineStochastic2018} analyzed variations of the gradient descent optimization algorithm, which approximately recover the correct computation result. Other ideas for approximate recovery are outlier detection methods based on pairwise distances between the computed gradient at different workers~\cite{blanchardMachineLearningAdversaries,chenDistributedStatisticalMachine2017,guerraoui2018hidden,el2020fast} or based on scoring algorithms~\cite{xieZenoDistributedStochastic}. These outlier detection methods can also be combined with redundancy-based methods~\cite{rajputDETOXRedundancybasedFramework,konstantinidisRobustDetection}.

Using such methods, however, the resulting gradient estimate may be only an inexact approximation of the desired total gradient in the error-free case. This increases the runtime for the gradient descent algorithm, see e.g.~\cite{bitarStochasticGradientCoding2020}, and references therein, and might perform poorly in some particular settings, e.g., when the distribution of the training data is not identical among the workers~\cite{chenRevisitingDistributedSynchronous2017,tandonGradientCoding2017}. Moreover, advanced gradient descent techniques, such as the momentum method~\cite{sutskeverImportanceInitializationMomentum}, are, in general, not compatible with approximate schemes~\cite{tandonGradientCoding2017}.
Due to the latter, the problem of tolerating Byzantine errors in distributed gradient descent with exact recovery has been approached within a coding-theoretic framework, termed DRACO~\cite{chenDRACOByzantineresilientDistributed2018}. In this framework, each gradient computation is replicated to several workers, and the workers send linear combinations of the computed gradients such that the main node can decode the exact gradient aggregate.
The authors of DRACO build on the idea of gradient coding~\cite{tandonGradientCoding2017}, which was originally designed to mitigate the effect of slow or unresponsive workers, commonly referred to as stragglers~\cite{DB13}. By replicating each gradient computation to $\nmalicious+1$ different workers, gradient coding can tolerate $\nmalicious$ stragglers by treating their responses as erasures. Applying the same ideas, DRACO can tolerate $\nmalicious$ malicious workers instead, treating the responses of malicious workers as errors in a codeword of an equivalent error correcting code. This comes at the cost of increasing the replication of each gradient computation to $2\nmalicious+1$, hence causing a large computation and communication overhead. Both DRACO and gradient coding were shown to achieve an optimal replication for the respective problem settings.

In this work, we consider the problem of exact gradient coding in the presence of an adversary controlling $\nmalicious$ malicious workers, which introduce Byzantine errors in their responses. In contrast to~\cite{chenDRACOByzantineresilientDistributed2018}, we define a more general framework in which the \master can run a small number of gradient computations itself to aid in decoding. We present a scheme that requires a replication of only $\nmalicious+1$ at the expense of running $\nmalicious$ local gradient computations. This scheme uses additional light communication between the workers and the \master to help identify which gradients should be computed locally. The ideas are extended to not only restrict to the extreme case of replication $\nmalicious+1$, but show how to leverage a higher replication $\nmalicious+\nhonest$ for $1 \leq \nhonest \leq \nmalicious$.
This higher replication can be exploited for complimentary tolerance of up to $\nhonest-1$ stragglers or alternatively to reduce the number of local computations at the main node by a factor of $1/\nhonest$. 
We augment our study with an intuitive high-level example of our ideas and provide a thorough discussion of our approach.

The idea of running extra computations at the \master has been exploited in former works in two different shapes. In~\cite{caoDistributedGradientDescent2019,prakash2020secure}, additional coarse gradients are computed at the \master to perform outlier detection in distributed gradient descent. In~\cite{soleymaniListDecodable2021}, local computations provide side information for pruning after list decoding. While the latter work is restricted to polynomial computations, the former focuses on approximate recovery, which is in contrast to our goals as outlined before.

\paragraph*{Organization}
We first present the problem of synchronous distributed gradient descent with Byzantine errors in \cref{sec:setting}. Then, we give an overview of our main results in \cref{sec:main_results}. Before we explain the results from a technical point of view, we give a motivating example to build an intuition for the problem in \cref{sec:example}. In \cref{sec:framework}, we formally define our framework. Finally, we provide the technical proofs of the fundamental limits in \cref{sec:fundamental_limits} before we propose our scheme and analyze its performance in \cref{sec:scheme}. In \cref{sec:discussion}, we discuss the scheme and compare it to the fundamental limits derived in the previous sections before \cref{sec:conclusion} concludes our work.

%% file: Journal/sections/problem_setting.tex
\section{Problem Setting}
\label{sec:setting}

\paragraph*{Notation}
Matrices and vectors are denoted by upper-case and lower-case bold letters, respectively. $\mathbf{A}_{i,j}$ refers to the element in row $i$ and column $j$ of the matrix $\mathbf{A}$. Scalars are denoted by lower-case letters, sets by calligraphic letters, and lists by fractal letters, respectively, e.g., $a$, $\mathcal{A}$ and $\lis{A}$. For an integer $a \geq 1$, we define $\range{a} \defeq \left\{ 1,2,\dots,a \right\}$.
Let $\lis{A}_i, \, i=1,\dots,t$, be a collection of lists, we define $\lis{A}^{(t)}$ to be their concatenation.
We use $\boldsymbol{1}_{m \times n}$ and $\boldsymbol{0}_{m \times n}$ to denote the all-one and all-zero matrices of dimension $m \times n$.

\vspace{1em}

\begin{figure}
    \centering
    \ifarxiv
    \resizebox{0.9\linewidth}{!}{ \input{Journal/tikz/main_worker} }
    \else
    \resizebox{0.6\linewidth}{!}{ \input{Journal/tikz/main_worker} }
    \fi
    \caption{Illustration of a distributed gradient descent setting with adversaries.}
\end{figure}

We consider a \emph{synchronous distributed gradient descent} setting, in which the goal is to fit the parameters $\params \in \R^\graddim$ of a model to a dataset consisting of $\ngrad$ samples $\sample \in \R^\graddim$, $\gind \in
\range{\ngrad}$. This is done by finding (local) optima for the problem
$\displaystyle \argmin_{\params \in \R^\graddim} \sum_{\gind \in \range{\ngrad}} \loss(\params, \sample)$ 
for a per sample loss function $\loss(\params, \sample)$.
The gradient descent algorithm starts with a random initialization for the 
parameter vector, defined as $\params^{(0)}$, and then iteratively applies the update rule $
    \params^{(\gditer+1)} = \params^{(\gditer)} - \frac{\learningrate}{\ngrad} \sum_{\gind \in
    \range{\ngrad}} \nabla \loss(\params^{(\gditer)}, \sample),
$ where $\gditer$ is the iteration index and $\learningrate \in \R$ is a hyper-parameter referred to as the learning rate. For notational convenience, we define the evaluation of the gradient of the loss function at individual samples as $
    \pgrad^{(\gditer)} \defeq \nabla \loss(\params^{(\gditer)}, \sample)
$ and call them \emph{partial gradients}.
Since, in practice, data is quantized to a finite set of values, we take the partial gradients to be vectors over a finite alphabet $\galpha$, i.e., $\pgrad^{(\gditer)} \in \galpha^\graddim$.

Consider a system comprising a \master and $\nworker$ worker nodes, $\nmalicious$ of which might be malicious.
The malicious workers can send arbitrarily corrupted information to the \master. At the start of the procedure, the \master distributes the samples to the workers with some redundancy.
Then, each iteration $\gditer$ starts with the \master broadcasting the current parameter vector $\params^{(\gditer)}$. The workers then compute the partial gradients $\pgrad^{(\gditer)}$ corresponding to the samples they store. At the end of the iteration, the \master must obtain the \emph{full gradient} $\tgrad^{(\gditer)} \defeq \sum_{\gind \in \range{\ngrad}} \pgrad^{(\gditer)}$ irrespective of the actions of the $\nmalicious$ malicious workers.

In this work, we are concerned with the problem of reliably reconstructing $\tgrad^{(\gditer)}$ \emph{exactly} at the \master in each iteration $\gditer$.
In the sequel, we consider a single iteration of gradient descent and omit the superscript $\gditer$.

%% file: Journal/tikz/main_worker.tex
\begin{tikzpicture}[every text node part/.style={align=center}]

    \def\myx{5*0.7}
    \def\myy{6*0.9}
    \def\mydx{0.3*0.9}
    \def\belowserver{2.0*0.9}

    \node[inner sep=0, font=\large] (s1) at (0,0)
        {\includegraphics[height=2cm]{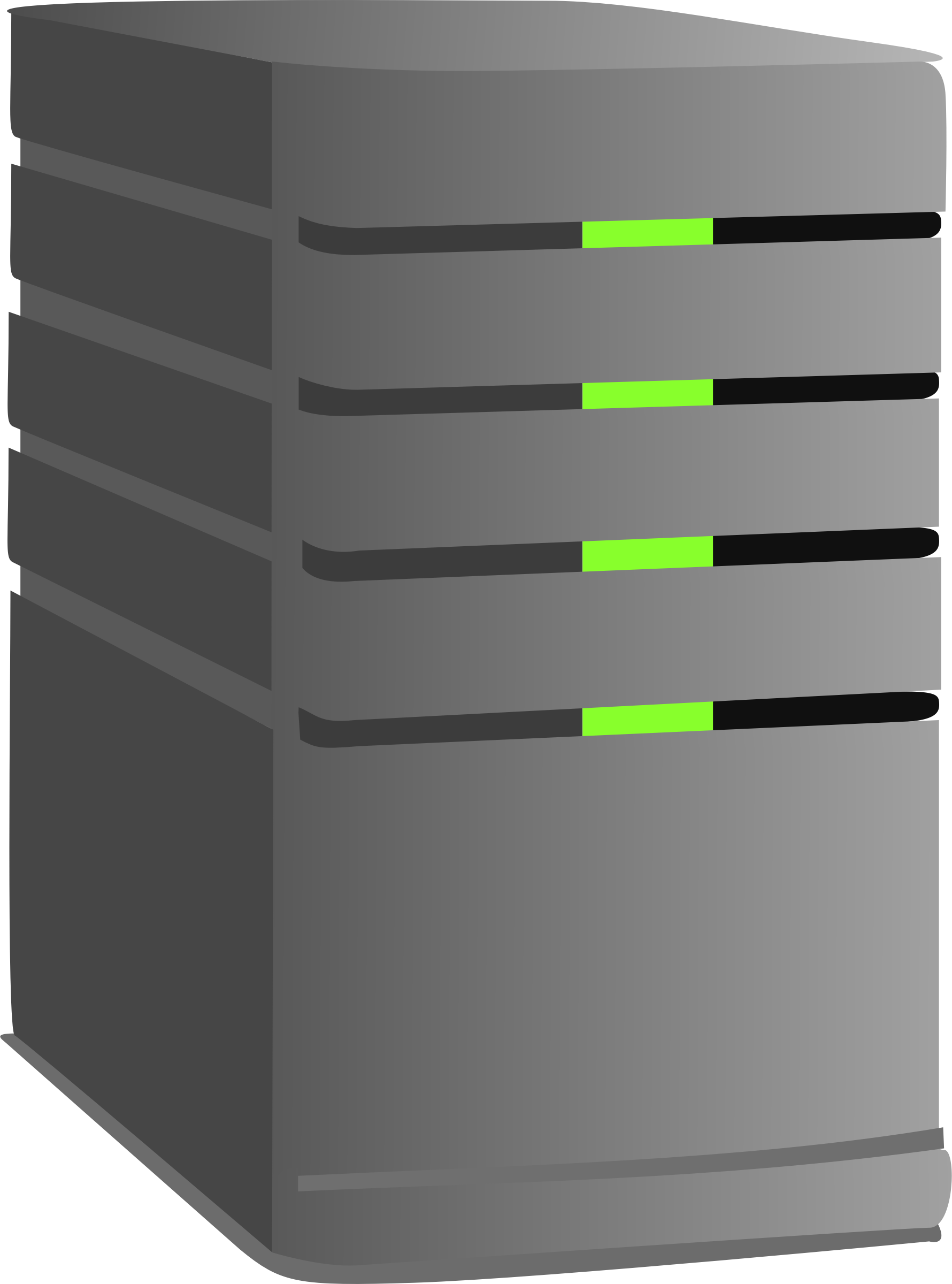}};
    \node[font=\large] (dataset) at (-2.5, 0.5) {$\mathbf{x}_1, \dots, \mathbf{x}_\ngrad$};

    \node[inner sep=0, font=\large] (us1) at (-1.5*\myx, -\myy) {\includegraphics[width=2cm]{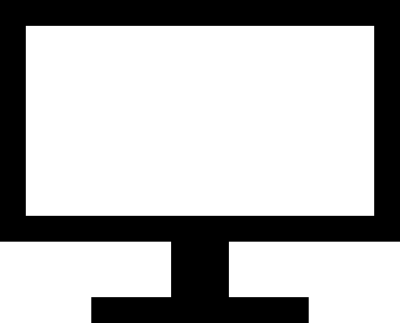}};
    \node[inner sep=0, font=\large] (us2) at (-0.5*\myx, -\myy) {\includegraphics[width=2cm]{Journal/pictures/monitor.png}};
    \node[inner sep=0, font=\large] (us4) at ( 0.5*\myx, -\myy) {\includegraphics[width=2cm]{Journal/pictures/monitor.png}};
    \node[inner sep=0, font=\large] (us5) at ( 1.5*\myx, -\myy) {\includegraphics[width=2cm]{Journal/pictures/monitor.png}};

   \node[inner sep=0, font=\large] (d1) at (-1.5*\myx, -1.2*\myy)
     {$\mathbf{x}_1, \mathbf{x}_2$};
   \node[inner sep=0, font=\large] (d2) at (-0.5*\myx, -1.2*\myy)
     {$\mathbf{x}_1, \mathbf{x}_2$};
   \node[inner sep=0, font=\large] (d4) at (1.5*\myx, -1.2*\myy)
     {$\mathbf{x}_{\ngrad-1}, \mathbf{x}_{\ngrad}$};

   \node[inner sep=0, font=\large] (w1) at (-1.5*\myx, -1.4*\myy)
     {worker node 1 \\ \phantom{malicious}};
   \node[inner sep=0, font=\large] (w2) at (-0.5*\myx, -1.4*\myy)
     {worker node 2 \\ (malicious)};
   \node[inner sep=0, font=\large] (w3) at (0.5*\myx, -1.4*\myy)
     {\dots \\ \phantom{(malicious)}};
   \node[inner sep=0, font=\large] (w4) at (1.5*\myx, -1.4*\myy)
     {worker node \nworker \\ \phantom{malicious}};

   \node[inner sep=0, font=\large] (mn) at (0, 0.3*\myy)
     {main node};

   \node[inner sep=0] (privacy) at (-0.5*\myx, -0.95*\myy)
        {\includegraphics[width=1cm]{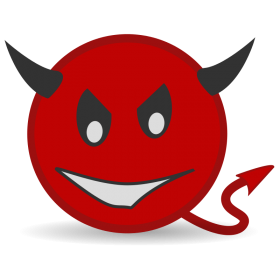}};

    \draw[->,line width=0.2mm] ({-1.5*(\myx-  \mydx)}, -0.75*\myy)--({-0.75*(\myx-2*\mydx)}, -\belowserver) 
    node [font=\large, pos=0.1, right] {$\mathrm{enc}_1(\mathbf{g}_1,
    \mathbf{g}_2)$};
    \draw[->,line width=0.2mm] ({-0.5*(\myx-2*\mydx)}, -0.75*\myy)--({-0.25*(\myx-4*\mydx)}, -\belowserver)
    node [font=\large, pos=0.1, right] {erroneous\\result};
    \draw[->,line width=0.2mm] ({ 0.5*(\myx+2*\mydx)}, -0.75*\myy)--({ 0.25*(\myx+4*\mydx)}, -\belowserver)
    node [font=\large, pos=0.1, right] {\dots};
    \draw[->,line width=0.2mm] ({ 1.5*(\myx+  \mydx)}, -0.75*\myy)--({ 0.75*(\myx+2*\mydx)}, -\belowserver)
    node [font=\large, pos=0.1, right] {$\mathrm{enc}_\nworker(\mathbf{g}_{\ngrad-1},
    \mathbf{g}_{\ngrad})$};

\end{tikzpicture}

%% file: Journal/sections/toy_example_extended.tex
\section{Motivating Example}
\label{sec:example}
Before formally stating the framework, the following example demonstrates the abstract problem and builds an intuition for our main results.
Consider a game among friends, \playA, \playB, and \playC, who play against \playboss.
\begin{enumerate}
    \item \playA, \playB, and \playC secretly agree on a list of eight integers $g_1, \dots, g_8$ and put them in separate envelopes.
    \item Two players from \playA, \playB, and \playC are designated as liars, without \playboss knowing which is which. The remaining player has to be truthful.
    \item \playboss's goal is to find the sum of the numbers $g_1 + \dots + g_8$. The game is played in rounds. In each round \playbosspronoun can first ask the other players questions about $g_1, \dots, g_8$ and then look in any number of envelopes.
    \item \playboss needs to find the correct sum every time. 
\end{enumerate}
What is the minimum number of envelopes \playboss needs to look inside?
If \playbosspronoun checks the minimum number of envelopes, how many questions does \playbosspronoun need to ask? 

If there was no limit on the number of questions, \playbosspronoun could ask each player for all values $g_1, \dots, g_8$. With $\exwpgrad{\gind}{\wind}$ we denote the value player $\wind\in \{\text{\playA}, \text{\playB}, \text{\playC}\}$ claims is in the envelope for $g_\gind$, $\gind \in [8]$.
Since the players only answer questions about these numbers, and assuming the liars are smart enough to avoid contradictions so as not to be detected, \playboss cannot gain any more information by asking more questions.

Not in every case can \playboss identify the correct sum based just on these answers.
For example, \cref{tab:example_key_error_pattern} shows three cases that are indistinguishable based on the values of all $\exwpgrad{\gind}{\wind}$.
\begin{table}[h]
    \caption{Example for a game with $2$ liars out of $3$ players. It shows a choice of values for which it is impossible to distinguish between the given cases.}
    \begin{minipage}{0.45\columnwidth}
    \begin{tabularx}{\columnwidth}{l*{5}{X}}
        & $\exwpgrad{1}{\wind}$ & $\exwpgrad{2}{\wind}$ & $\exwpgrad{3}{\wind}$ & \dots & $\exwpgrad{8}{\wind}$ \\
        \toprule
        \playA & $1$ & $3$ & $4$ & \dots & $4$ \\
        \playB & $2$ & $7$ & $4$ & \dots & $4$ \\
        \playC & $2$ & $3$ & $4$ & \dots & $4$ \\
    \end{tabularx}
    \end{minipage}%
    \hfill
    \begin{minipage}{0.49\columnwidth}
        \begin{align*}
            \text{\emph{Case 1:  }} 
                            & g_1=2;\ g_2=3 \\
                            & \text{\playA and \playB lie} \\
                \text{\emph{Case 2:  }} 
                            & g_1=1;\ g_2=3 \\
                            & \text{\playB and \playC lie} \\
                \text{\emph{Case 3:  }}
                            & g_1=2;\ g_2=7 \\
                            & \text{\playA and \playC lie}
        \end{align*}
    \end{minipage}
    \label{tab:example_key_error_pattern}
\end{table}

To resolve the conflict, \playboss chooses one of the integers $g_i$, $i\in[8]$ on which at least two players' answers disagree and checks the corresponding envelope. \playbossPronoun is guaranteed to identify at least one liar. After eliminating the identified liar(s), \playbosspronoun repeats the process.
After opening at most two envelopes, it is guaranteed that all non-eliminated players, including the honest one, agree on the sum $g_1 + \dots + g_8$, and it is therefore correct.
It can be verified that the cases in \cref{tab:example_key_error_pattern} cannot be distinguished after opening any one envelope. Thus, using this strategy \playboss opens the smallest possible number of envelopes in the worst case.

Now, \playboss additionally wishes to minimize the number of questions asked.
With the above strategy, \playboss asks each player $8$ questions, which is $24$ questions in total.
\playboss realizes that using multiple rounds, \playbosspronoun can reduce the number of questions \playbosspronoun needs to ask using the following recursive procedure.
First, \playbosspronoun asks every player for the desired total sum $g_1 + \dots + g_8$ and the sum of the first half of the values, $g_1 + g_2 + g_3 + g_4$.
\playboss picks any two players whose values for the total sum differ.
\playbossPronoun infers their values for $g_5 + \dots + g_8$ as $(g_1 + \dots + g_8) - (g_1 + \dots + g_4)$.
Since the two players disagree on the total sum, they are guaranteed to disagree on either $g_1 + \dots + g_4$ or $g_5 + \dots + g_8$.
Assume they disagree on $g_5 + \dots + g_8$, then \playboss requests $g_5 + g_6$ from both players.
Again, they are guaranteed to disagree either on $g_5 + g_6$ or $g_7 + g_8$.
Assume they disagree on $g_5+g_6$. \playboss asks for $g_5$.
After having obtained a single integer, either $g_5$ or $g_6$, on which the two players disagree, \playboss opens the corresponding envelope and thus identifies and eliminates at least one liar.
In the worst case \playboss has to repeat the above procedure once more on the remaining two players. Note that in the second match, \playboss already knows the sum $g_1 + g_2 + g_3 + g_4$ and $g_5 + g_6 + g_7 + g_8$ for one player from the first match. Therefore, \playbosspronoun asks only $14$ questions as opposed to the $24$ from before.

This game is the crux of our framework. The private numbers are the partial gradients computed at the workers, opening an envelope represents a local computation at the \master, and asking questions is the light communication between the \master and the workers. The figures of merit of this game are:
\begin{enumerate*}[label=\arabic*)]
        \item the minimum number of envelopes that \playbossletter needs to open; and 
        \item  the number of questions that \playbossletter needs to ask the other players to recover the desired sum correctly.
\end{enumerate*}

In our work, we also generalize the above ideas to cases where there are several honest worker. We give an intuition for this case by extending the above example: consider the same setting again but with an additional honest player \playD, and suppose \playC is honest, too. Although \playboss does not know the honest players' identities, \playbosspronoun knows that at least two players respond honestly and, thus, agree in their responses to the same question. That is, if a player's response is not supported by a second player, \playbosspronoun can expose a player as a liar without having to open an envelope. Having this in mind, the liars might decide to cooperate and align their responses. In this case, however, if \playboss employs one of the above strategies and finds an integer on which some players disagree, \playbosspronoun will identify both liars at the same time by opening a single envelope. A detailed explanation of this example is given in \cref{subsec:general_construction}.

%% file: Journal/sections/main_results.tex
\section{Main Results}
\label{sec:main_results}

In order to solve the problem of Byzantine errors in synchronous distributed gradient descent, we propose a novel framework for Byzantine-resilient gradient coding in the presence of $\nmalicious$ malicious workers. This framework augments the previous ideas of gradient coding~\cite{tandonGradientCoding2017,chenDRACOByzantineresilientDistributed2018} with an interactive communication protocol and a small number of local computations at the \master, and is formally defined in \cref{sec:framework}. We refer to this framework as \BGC schemes and characterize schemes by a tuple of parameters $(\nround,\localcomp,\replfact,\commoh)$, which are explained in the following. Our main result consists of fundamental limits of the novel framework in terms of the following figures of merit (informal):
\begin{itemize}
    \item The {\bf replication factor} $\replfact$ is the average number of workers to which each sample is assigned.
    \item The {\bf communication overhead} $\commoh$ is the maximum number of bits transmitted from each worker to the main node during the interactive communication protocol.
    \item The {\bf computation overhead} $\localcomp$ is the maximum number of partial gradients that the \master needs to compute locally.
\end{itemize}
Furthermore, we propose an achievable scheme and analyze its optimality with regard to our figures of merit. In the sequel, we summarize our main results. We focus on a fractional repetition data assignment, which is defined in \cref{sec:framework}.

\vspace{1em}
\paragraph{Trade-off between replication and computation overhead}
We first present a fundamental trade-off between the replication factor $\replfact$ and the number of local computations $\localcomp$ at the \master. For the non-trivial case of $\localcomp<\ngrad$, i.e., when the \master does not compute all the partial gradients locally, the replication factor of any \BGC scheme is bounded from below by $\replfact \geq {\nmalicious+1}$.
In addition, we note that if $\localcomp = 0$, then the replication factor of any \BGC scheme is bounded from below as $\replfact \geq {2\nmalicious+1}$ and can be achieved through DRACO \cite{chenDRACOByzantineresilientDistributed2018} with $\localcomp=\commoh=0$. Thus, we focus on the case $\nmalicious+1 \leq \replfact \leq 2\nmalicious+1$, and we investigate the fundamental tradeoff between $\localcomp$, $\replfact$, and $\commoh$. In particular, we show that for $\localcomp = \nmalicious$ the replication factor $\replfact = {\nmalicious+1}$ is achievable. For $0 \leq \localcomp<\nmalicious$, we show that for any $1 \leq \nhonest \leq \nmalicious+1$, if $\replfact \leq {\nmalicious + \nhonest}$, then $\localcomp\geq \compbound$. This result is stated formally in \cref{thm:converse_C}. \cref{fig:converse_C} visualized the trade-off between the number of local computations at the \master and the replication factor.

\begin{theorem}[Lower bound on $\localcomp$ and $\replfact$]
Suppose that \mbox{$\nworker = \ngroup (\nmalicious + \nhonest)$} for integers $\ngroup,\nhonest \geq 1$. For any \BGC scheme with parameters (\nround,\localcomp,\replfact,\commoh) that has a fractional repetition data assignment, it holds that if $\replfact \leq \nmalicious + \nhonest$, then $\localcomp\geq \compbound$ (conversely,  if $\localcomp < \compbound$ then $\replfact > {\nmalicious + \nhonest}$).
\label{thm:converse_C}
\end{theorem}

\begin{figure}[h]
    \centering
      \resizebox{\linewidth}{!}{\input{tikz/c_vs_rhorel.tex}}
    \caption{Trade-off between local computations $\localcomp$ and (normalized) replication $\bar{\replfact}=\replfact/\nmalicious$.  For $\bar{\replfact} \geq 1 + \frac{1}{s}$, points on or above the blue curve $c = \lfloor \frac{1}{\bar{\replfact}-1} \rfloor$ are achievable. Points below or to the left of the curve are fundamentally impossible, including the shaded region $\bar{\replfact} \leq 1$, which is unattainable for any number of local computations.}
    \label{fig:converse_C}
\end{figure}

\paragraph{Lower bound on the communication overhead}
Second, we analyze limits on the communication overhead $\commoh$ for schemes that achieve the minimal number of local computations. We prove a fundamental lower bound on the communcation overhead, as state in~\cref{thm:converse_commoh}.

\begin{theorem}[Lower bound on $\commoh$ for fixed $\localcomp$ and $\replfact$]
Suppose that $\nworker = \ngroup (\nmalicious + \nhonest)$ for integers $\ngroup,\nhonest \geq 1$.
For any \BGC scheme with parameters (\nround,\localcomp,\replfact,\commoh) that has a fractional repetition data assignment with $\replfact = {\nmalicious + \nhonest}$ and $\localcomp = \compbound$, it holds that
   \begin{align*}
       \commoh \geq {\log_{2} \binom{\ngrad/\ngroup}{\lfloor \nmalicious / \nhonest \rfloor}}.%
   \end{align*}
   \label{thm:converse_commoh}
   \vspace{-1em}
\end{theorem}

\begin{figure}[ht]
    \centering
        \resizebox{0.9\linewidth}{!}{\input{tikz/kappa_vs_p_updated2}}
    \caption{Comparison of converse and achievability for $\commoh$ over the dataset size $\ngrad$. We consider a system of $\nworker=10$ workers, $\ngroup=1$ group and an alphabet size $|\galpha|=2^{16}$. As the percentage of malicious workers rises from \SI{50}{\percent} to \SI{90}{\percent} the communication overhead of the scheme as well as the lower bound increase.
    }
    \label{fig:achievability_vs_converse_datacolumns}
\end{figure}
\paragraph{Byzantine-resilient gradient coding scheme}
We finally present a scheme that allows the reconstruction of the full gradient in the presence of $\nmalicious$ malicious workers by replicating each partial gradient only $\nmalicious+1$ times. We show that the scheme achieves the optimal trade-off between the replication factor $\replfact$ and the number of local computations $\localcomp$ at the \master. The communication overhead is a constant factor away from the lower bound.
The capabilities of our scheme are summarized in \cref{thm:scheme}. A comparison of the communication overhead of the scheme and the bound in~\cref{thm:converse_commoh} can be found in~\cref{fig:achievability_vs_converse_datacolumns}.
For ease of notation let $\bar{\localcomp} = \max\{1,\localcomp\}$.
\begin{theorem}
  The scheme constructed in \cref{sec:scheme} is an \BGC scheme with parameters (\nround,\localcomp,\replfact,\commoh), and for any $\nhonest$, $1 \leq \nhonest \leq \nmalicious+1$ achieves
    \begin{align*}
      \nround \leq&
      \left( \nmalicious - \bar{\localcomp} (\nhonest-1) \right)
      \left( 2\left\lceil \log_2\left(\frac{\ngrad}{\ngroup}\right) \right\rceil + 1 \right),\\
      \localcomp \leq& \compbound,\\
      \replfact =& {\nmalicious+\nhonest},\\
      \commoh \leq&
      \left( \nmalicious-\bar{\localcomp}(\nhonest-1) \right)
      \cdot \Big(
        \left( 1 + \left\lceil \log_2{\card{\galpha}} \right\rceil \right) \left\lceil \log_2\left( \frac{\ngrad}{\ngroup}\right) \right\rceil \\
                  &+ \frac{\nmalicious+(\bar{\localcomp}+2)\nhonest-3}{2} \Big) - \bar{\localcomp} \frac{\nmalicious-\nhonest + 1}{2}.
    \end{align*}
  \label{thm:scheme}
  \vspace{-2em}
\end{theorem}
The following corollary states a simpler equation for the communication overhead for the case of a large dataset size. 
\begin{corollary}
  \label{cor:achievable_commoh}
  For large dataset sizes $\ngrad$, the scheme constructed in \cref{sec:scheme} achieves an asymptotic communication overhead of
  \vspace{-0.5em}
  \begin{equation*}
    \commoh \leq
    \left( \nmalicious-\bar{\localcomp}(\nhonest-1) \right)
    \left( 1+ \left\lceil \log_2{\card{\galpha}} \right\rceil \right)
    \left\lceil \log_2\left( \frac{\ngrad}{\ngroup}\right) \right\rceil.
  \end{equation*}
  \vspace{-0.5em}
  In particular, if $\localcomp=\compbound$ and $\nhonest \mid \nmalicious$,
  \begin{equation*}
    \commoh \leq
    \compbound
    \left( 1+ \left\lceil \log_2{\card{\galpha}} \right\rceil \right)
    \left\lceil \log_2\left( \frac{\ngrad}{\ngroup}\right) \right\rceil.
  \end{equation*}
\end{corollary}
The fundamental limits established in \cref{thm:converse_C} and \cref{thm:converse_commoh}, as well as the parameters of our \BGC scheme given in \cref{thm:scheme} can be extended to scenarios with stragglers. The results are summarized in the following corollary.
\begin{corollary}
  \label{cor:stagglers}
  Consider any \BGC scheme with parameters (\nround,\localcomp,\replfact,\commoh) that has a fractional repetition data assignment with $\replfact = \nmalicious+\nhonest$.
  In the presence of $\nmalicious$ malicious workers and $\nstraggle$ stragglers with $0 \leq \nstraggle < \nhonest$, the \BGC scheme must have $\localcomp \geq \left\lfloor \frac{\nmalicious}{\nhonest-\nstraggle} \right\rfloor$.
  If it achieves the minimal number of local computations, then
  $\commoh \geq {\log_{2} \binom{\ngrad/\ngroup}{\lfloor \nmalicious / (\nhonest-\nstraggle) \rfloor}}$.
  The \BGC scheme constructed in \cref{sec:scheme} achieves the minimal number of local computations and an asymptotic communication overhead of
    \begin{align*}
      \commoh \leq &
      \left( \nmalicious-\bar{\localcomp}(\nhonest-\nstraggle-1) \right) \cdot \\
                   &\left( 1+ \left\lceil \log_2{\card{\galpha}} \right\rceil \right) 
                   \left\lceil \log_2\left( \frac{\ngrad}{\ngroup}\right) \right\rceil.
    \end{align*}
\end{corollary}

%% file: tikz/c_vs_rhorel.tex
\begin{tikzpicture}

\begin{axis}[%
scale only axis,
width=100mm,
height=50mm,
xmin=0,
xmax=3,
ymin=0,
ymax=10,
ylabel={local computations $\localcomp$},
xlabel={normalized replication $\bar{\replfact}$},
xtick={0,1,2}
]
\addplot [color={rgb,1:red,0.0667;green,0.4392;blue,0.6667},line width={1}]
  table[row sep=crcr]{%
1.01	99\\
1.02	49\\
1.03	33\\
1.04	24\\
1.05	19\\
1.06	16\\
1.07	14\\
1.08	12\\
1.09	11\\
1.1	9\\
1.11	9\\
1.12	8\\
1.13	7\\
1.14	7\\
1.15	6\\
1.16	6\\
1.17	5\\
1.18	5\\
1.19	5\\
1.2	5\\
1.21	4\\
1.22	4\\
1.23	4\\
1.24	4\\
1.25	4\\
1.26	3\\
1.27	3\\
1.28	3\\
1.29	3\\
1.3	3\\
1.31	3\\
1.32	3\\
1.33	3\\
1.34	2\\
1.35	2\\
1.36	2\\
1.37	2\\
1.38	2\\
1.39	2\\
1.4	2\\
1.41	2\\
1.42	2\\
1.43	2\\
1.44	2\\
1.45	2\\
1.46	2\\
1.47	2\\
1.48	2\\
1.49	2\\
1.5	2\\
1.51	1\\
1.52	1\\
1.53	1\\
1.54	1\\
1.55	1\\
1.56	1\\
1.57	1\\
1.58	1\\
1.59	1\\
1.6	1\\
1.61	1\\
1.62	1\\
1.63	1\\
1.64	1\\
1.65	1\\
1.66	1\\
1.67	1\\
1.68	1\\
1.69	1\\
1.7	1\\
1.71	1\\
1.72	1\\
1.73	1\\
1.74	1\\
1.75	1\\
1.76	1\\
1.77	1\\
1.78	1\\
1.79	1\\
1.8	1\\
1.81	1\\
1.82	1\\
1.83	1\\
1.84	1\\
1.85	1\\
1.86	1\\
1.87	1\\
1.88	1\\
1.89	1\\
1.9	1\\
1.91	1\\
1.92	1\\
1.93	1\\
1.94	1\\
1.95	1\\
1.96	1\\
1.97	1\\
1.98	1\\
1.99	1\\
2	1\\
2.01	0\\
};
\addplot [color={rgb,1:red,0.0667;green,0.4392;blue,0.6667},line width={4}]
  table[row sep=crcr]{%
2.01	0\\
2.02	0\\
2.03	0\\
2.04	0\\
2.05	0\\
2.06	0\\
2.07	0\\
2.08	0\\
2.09	0\\
2.1	0\\
2.11	0\\
2.12	0\\
2.13	0\\
2.14	0\\
2.15	0\\
2.16	0\\
2.17	0\\
2.18	0\\
2.19	0\\
2.2	0\\
2.21	0\\
2.22	0\\
2.23	0\\
2.24	0\\
2.25	0\\
2.26	0\\
2.27	0\\
2.28	0\\
2.29	0\\
2.3	0\\
2.31	0\\
2.32	0\\
2.33	0\\
2.34	0\\
2.35	0\\
2.36	0\\
2.37	0\\
2.38	0\\
2.39	0\\
2.4	0\\
2.41	0\\
2.42	0\\
2.43	0\\
2.44	0\\
2.45	0\\
2.46	0\\
2.47	0\\
2.48	0\\
2.49	0\\
2.5	0\\
2.51	0\\
2.52	0\\
2.53	0\\
2.54	0\\
2.55	0\\
2.56	0\\
2.57	0\\
2.58	0\\
2.59	0\\
2.6	0\\
2.61	0\\
2.62	0\\
2.63	0\\
2.64	0\\
2.65	0\\
2.66	0\\
2.67	0\\
2.68	0\\
2.69	0\\
2.7	0\\
2.71	0\\
2.72	0\\
2.73	0\\
2.74	0\\
2.75	0\\
2.76	0\\
2.77	0\\
2.78	0\\
2.79	0\\
2.8	0\\
2.81	0\\
2.82	0\\
2.83	0\\
2.84	0\\
2.85	0\\
2.86	0\\
2.87	0\\
2.88	0\\
2.89	0\\
2.9	0\\
2.91	0\\
2.92	0\\
2.93	0\\
2.94	0\\
2.95	0\\
2.96	0\\
2.97	0\\
2.98	0\\
2.99	0\\
3	0\\
};

\addplot[color=black, draw=none, name path=zero] coordinates {(0, 0) (0, 10)};
\addplot[color=black, dashed, name path=one] coordinates {(1, 0) (1, 10)};
\addplot[color=black, dashed, name path=two] coordinates {(2, 0) (2, 10)};

\pgfdeclarepatternformonly{north east lines wide}%
    {\pgfqpoint{-1pt}{-1pt}}%
    {\pgfqpoint{10pt}{10pt}}%
    {\pgfqpoint{9pt}{9pt}}%
    {
      \pgfsetlinewidth{0.4pt}
      \pgfpathmoveto{\pgfqpoint{0pt}{0pt}}
      \pgfpathlineto{\pgfqpoint{9.1pt}{9.1pt}}
      \pgfusepath{stroke}
     }
\tikzfillbetween[of=zero and one, on layer=standard] {pattern=north east lines wide, opacity=0.4};

\end{axis}
\end{tikzpicture}%

%% file: tikz/kappa_vs_p_updated2.tex
\begin{tikzpicture}
\begin{axis}[
  point meta max={nan}, 
  point meta min={nan}, 
  legend cell align={left}, 
  title={}, 
  title style={at={{(0.5,1)}}, anchor={south}, font={{\fontsize{14 pt}{18.2 pt}\selectfont}}, color={rgb,1:red,0.0;green,0.0;blue,0.0}, draw opacity={1.0}, rotate={0.0}}, 
  legend columns=3, 
  legend style={
            /tikz/column 2/.style={
                column sep=3pt,
            },
      font={{\fontsize{7 pt}{10.4 pt}\selectfont}},
      cells={anchor={center}}, 
      at={(0.11,1.05)},
      anchor=south west
      }, 
  axis background/.style={fill={rgb,1:red,1.0;green,1.0;blue,1.0}, opacity={1.0}}, 
  anchor={north west}, 
  xshift={1.0mm}, 
  yshift={-1.0mm}, 
width={100mm}, 
height={50mm}, 
  xlabel={number of samples $\ngrad$}, 
  x tick style={color={rgb,1:red,0.0;green,0.0;blue,0.0}, opacity={1.0}}, 
  x tick label style={color={rgb,1:red,0.0;green,0.0;blue,0.0}, opacity={1.0}, rotate={0}}, 
  xmode={log}, 
  log basis x={10}, 
  xmajorgrids={true}, 
  xmin={100}, 
  xmax={1.3182567385564074e6}, 
  xtick={{100.0,1000.0,10000.0,100000.0,1.0e6}}, 
  xticklabels={{$10^{2}$,$10^{3}$,$10^{4}$,$10^{5}$,$10^{6}$}}, 
  xtick align={inside}, 
  xticklabel style={font={{\fontsize{8 pt}{10.4 pt}\selectfont}}, color={rgb,1:red,0.0;green,0.0;blue,0.0}, draw opacity={1.0}, rotate={0.0}}, 
  x grid style={color={rgb,1:red,0.0;green,0.0;blue,0.0}, draw opacity={0.1}, line width={0.5}, solid}, 
  axis x line*={left}, 
  x axis line style={color={rgb,1:red,0.0;green,0.0;blue,0.0}, draw opacity={1.0}, line width={1}, solid}, 
  ylabel={communication overhead $\kappa$}, 
  y tick style={color={rgb,1:red,0.0;green,0.0;blue,0.0}, opacity={1.0}}, 
  y tick label style={color={rgb,1:red,0.0;green,0.0;blue,0.0}, opacity={1.0}, rotate={0}}, 
  ymode={log}, 
  log basis y={10}, 
  ymajorgrids={true}, 
  ymin={1}, 
  ymax={10000}, 
  ytick={{1.0,10.0,100.0,1000.0,10000.0}}, 
  yticklabels={{$10^{0}$,$10^{1}$,$10^{2}$,$10^{3}$,$10^{4}$}}, 
  ytick align={inside}, 
  yticklabel style={font={{\fontsize{8 pt}{10.4 pt}\selectfont}}, color={rgb,1:red,0.0;green,0.0;blue,0.0}, draw opacity={1.0}, rotate={0.0}}, 
  y grid style={color={rgb,1:red,0.0;green,0.0;blue,0.0}, draw opacity={0.1}, line width={0.5}, solid}, 
  axis y line*={left}, 
  y axis line style={color={rgb,1:red,0.0;green,0.0;blue,0.0}, draw opacity={1.0}, line width={1}, solid}, 
  colorbar={false}
  ]
  \addplot[color={rgb,1:red,0.0667;green,0.4392;blue,0.6667}, 
  name path={0e242e9c-d481-452f-a98b-d4d0f32d2b0d}, 
  draw opacity={1.0}, 
  line width={1},
  solid]
        table[row sep={\\}]
        {
            \\
            100.0  1107.0  \\
            126.0  1107.0  \\
            158.0  1260.0  \\
            200.0  1260.0  \\
            251.0  1260.0  \\
            316.0  1413.0  \\
            398.0  1413.0  \\
            501.0  1413.0  \\
            631.0  1566.0  \\
            794.0  1566.0  \\
            1000.0  1566.0  \\
            1259.0  1719.0  \\
            1585.0  1719.0  \\
            1995.0  1719.0  \\
            2512.0  1872.0  \\
            3162.0  1872.0  \\
            3981.0  1872.0  \\
            5012.0  2025.0  \\
            6310.0  2025.0  \\
            7943.0  2025.0  \\
            10000.0  2178.0  \\
            12589.0  2178.0  \\
            15849.0  2178.0  \\
            19953.0  2331.0  \\
            25119.0  2331.0  \\
            31623.0  2331.0  \\
            39811.0  2484.0  \\
            50119.0  2484.0  \\
            63096.0  2484.0  \\
            79433.0  2637.0  \\
            100000.0  2637.0  \\
            125893.0  2637.0  \\
            158489.0  2790.0  \\
            199526.0  2790.0  \\
            251189.0  2790.0  \\
            316228.0  2943.0  \\
            398107.0  2943.0  \\
            501187.0  2943.0  \\
            630957.0  3096.0  \\
            794328.0  3096.0  \\
            1.0e6  3096.0  \\
        }
        ;
    \addlegendentry {scheme $\nmalicious=9$}
    \addplot[color={rgb,1:red,0.6392;green,0.6745;blue,0.7255}, name path={5c6cc1ca-1ded-4112-8f2e-02f121141a9a}, draw opacity={1.0}, line width={1}, solid]
        table[row sep={\\}]
        {
            \\
            100.0  376.0  \\
            126.0  376.0  \\
            158.0  427.0  \\
            200.0  427.0  \\
            251.0  427.0  \\
            316.0  478.0  \\
            398.0  478.0  \\
            501.0  478.0  \\
            631.0  529.0  \\
            794.0  529.0  \\
            1000.0  529.0  \\
            1259.0  580.0  \\
            1585.0  580.0  \\
            1995.0  580.0  \\
            2512.0  631.0  \\
            3162.0  631.0  \\
            3981.0  631.0  \\
            5012.0  682.0  \\
            6310.0  682.0  \\
            7943.0  682.0  \\
            10000.0  733.0  \\
            12589.0  733.0  \\
            15849.0  733.0  \\
            19953.0  784.0  \\
            25119.0  784.0  \\
            31623.0  784.0  \\
            39811.0  835.0  \\
            50119.0  835.0  \\
            63096.0  835.0  \\
            79433.0  886.0  \\
            100000.0  886.0  \\
            125893.0  886.0  \\
            158489.0  937.0  \\
            199526.0  937.0  \\
            251189.0  937.0  \\
            316228.0  988.0  \\
            398107.0  988.0  \\
            501187.0  988.0  \\
            630957.0  1039.0  \\
            794328.0  1039.0  \\
            1.0e6  1039.0  \\
        }
        ;
    \addlegendentry {scheme $\nmalicious=7$}
    \addplot[color={rgb,1:red,0.9882;green,0.4902;blue,0.0431}, name path={dfcf77df-a2d1-4ea1-845c-33668df8d22a}, draw opacity={1.0}, line width={1}, solid]
        table[row sep={\\}]
        {
            \\
            100.0  127.0  \\
            126.0  127.0  \\
            158.0  144.0  \\
            200.0  144.0  \\
            251.0  144.0  \\
            316.0  161.0  \\
            398.0  161.0  \\
            501.0  161.0  \\
            631.0  178.0  \\
            794.0  178.0  \\
            1000.0  178.0  \\
            1259.0  195.0  \\
            1585.0  195.0  \\
            1995.0  195.0  \\
            2512.0  212.0  \\
            3162.0  212.0  \\
            3981.0  212.0  \\
            5012.0  229.0  \\
            6310.0  229.0  \\
            7943.0  229.0  \\
            10000.0  246.0  \\
            12589.0  246.0  \\
            15849.0  246.0  \\
            19953.0  263.0  \\
            25119.0  263.0  \\
            31623.0  263.0  \\
            39811.0  280.0  \\
            50119.0  280.0  \\
            63096.0  280.0  \\
            79433.0  297.0  \\
            100000.0  297.0  \\
            125893.0  297.0  \\
            158489.0  314.0  \\
            199526.0  314.0  \\
            251189.0  314.0  \\
            316228.0  331.0  \\
            398107.0  331.0  \\
            501187.0  331.0  \\
            630957.0  348.0  \\
            794328.0  348.0  \\
            1.0e6  348.0  \\
        }
        ;
    \addlegendentry {scheme $\nmalicious=5$}
    \addplot[color={rgb,1:red,0.0667;green,0.4392;blue,0.6667}, name path={916ac44c-4455-46cd-b337-0e0654368a75}, draw opacity={1.0}, line width={1}, dashed]
        table[row sep={\\}]
        {
            \\
            100.0  40.79083020422186  \\
            126.0  43.90459398717431  \\
            158.0  46.93012042802772  \\
            200.0  50.06212876806595  \\
            251.0  53.06546257833338  \\
            316.0  56.0990423054254  \\
            398.0  59.129054537888244  \\
            501.0  62.14461013031177  \\
            631.0  65.1616515804815  \\
            794.0  68.16217721787014  \\
            1000.0  71.170840739337  \\
            1259.0  74.17208434014132  \\
            1585.0  77.1704445359738  \\
            1995.0  80.16435361338267  \\
            2512.0  83.16175479455387  \\
            3162.0  86.15402174309172  \\
            3981.0  89.14804786498459  \\
            5012.0  92.1410348768562  \\
            6310.0  95.13345098392195  \\
            7943.0  98.12354052211488  \\
            10000.0  101.11508322179576  \\
            12589.0  104.10562507133712  \\
            15849.0  107.09652716120708  \\
            19953.0  110.08712614920758  \\
            25119.0  113.0772218025865  \\
            31623.0  116.06740402133529  \\
            39811.0  119.05747770792563  \\
            50119.0  122.04746072942883  \\
            63096.0  125.03739214640412  \\
            79433.0  128.02727095000495  \\
            100000.0  131.01711186517088  \\
            125893.0  134.00700129886349  \\
            158489.0  136.99674795962883  \\
            199526.0  139.98656173580966  \\
            251189.0  142.97638407061154  \\
            316228.0  145.9661530437652  \\
            398107.0  148.95590693901033  \\
            501187.0  151.94566856669417  \\
            630957.0  154.93542412930236  \\
            794328.0  157.9251795967479  \\
            1.0e6  160.91493216691936  \\
        }
        ;
    \addlegendentry {bound $\nmalicious=9$}
    \addplot[color={rgb,1:red,0.6392;green,0.6745;blue,0.7255}, name path={18869fbd-03e6-4fe7-a510-983c69edb319}, draw opacity={1.0}, line width={1}, dashed]
        table[row sep={\\}]
        {
            \\
            100.0  12.273212809854334  \\
            126.0  12.943064208162003  \\
            158.0  13.59840149706873  \\
            200.0  14.280480810318373  \\
            251.0  14.937327838612859  \\
            316.0  15.602988766564382  \\
            398.0  16.269619817686607  \\
            501.0  16.934451077857297  \\
            631.0  17.600704213369827  \\
            794.0  18.264172252846937  \\
            1000.0  18.930125152454504  \\
            1259.0  19.594978774598307  \\
            1585.0  20.259623745106378  \\
            1995.0  20.923622725507904  \\
            2512.0  21.588667062163125  \\
            3162.0  22.252786972262868  \\
            3981.0  22.917467871522362  \\
            5012.0  23.582053521751437  \\
            6310.0  24.246619925322847  \\
            7943.0  24.91075493348429  \\
            10000.0  25.575280482380855  \\
            12589.0  26.239637530802405  \\
            15849.0  26.904117360163006  \\
            19953.0  27.568563806139288  \\
            25119.0  28.232925385459655  \\
            31623.0  28.89732761525845  \\
            39811.0  29.561722741015988  \\
            50119.0  30.226111232530847  \\
            63096.0  30.89049898941795  \\
            79433.0  31.554883579708555  \\
            100000.0  32.21926652185108  \\
            125893.0  32.88366562401786  \\
            158489.0  33.54803727192452  \\
            199526.0  34.21242722744987  \\
            251189.0  34.876821780410175  \\
            316228.0  35.5412066165111  \\
            398107.0  36.2055898027131  \\
            501187.0  36.86997605813843  \\
            630957.0  37.53436203885235  \\
            794328.0  38.198748850789805  \\
            1.0e6  38.863135695952586  \\
        }
        ;
    \addlegendentry {bound $\nmalicious=7$}
    \addplot[color={rgb,1:red,0.9882;green,0.4902;blue,0.0431}, name path={a9e21dfe-363f-47d4-9a89-d46c55fb56ce}, draw opacity={1.0}, line width={1}, dashed]
        table[row sep={\\}]
        {
            \\
            100.0  6.643856189774724  \\
            126.0  6.977279923499917  \\
            158.0  7.303780748177103  \\
            200.0  7.643856189774724  \\
            251.0  7.971543553950772  \\
            316.0  8.303780748177102  \\
            398.0  8.63662462054365  \\
            501.0  8.968666793195208  \\
            631.0  9.30149619498255  \\
            794.0  9.632995197142957  \\
            1000.0  9.965784284662087  \\
            1259.0  10.298062567719017  \\
            1585.0  10.63026712502677  \\
            1995.0  10.962173031109709  \\
            2512.0  11.294620748891626  \\
            3162.0  11.62662165235837  \\
            3981.0  11.958915156091349  \\
            5012.0  12.29117069932186  \\
            6310.0  12.623424289869911  \\
            7943.0  12.955468287959821  \\
            10000.0  13.287712379549449  \\
            12589.0  13.61987606750398  \\
            15849.0  13.95210419527352  \\
            19953.0  14.284318056309619  \\
            25119.0  14.61649141050838  \\
            31623.0  14.948686618840247  \\
            39811.0  15.280879490037275  \\
            50119.0  15.613070009104804  \\
            63096.0  15.945260927339202  \\
            79433.0  16.277450871118504  \\
            100000.0  16.609640474436812  \\
            125893.0  16.94183854187783  \\
            158489.0  17.274023187380944  \\
            199526.0  17.60621722903987  \\
            251189.0  17.938413761942925  \\
            316228.0  18.270605589358617  \\
            398107.0  18.60279671330265  \\
            501187.0  18.934989468348846  \\
            630957.0  19.267182162686478  \\
            794328.0  19.599375333518466  \\
            1.0e6  19.931568569324174  \\
        }
        ;
    \addlegendentry {bound $\nmalicious=5$}
\end{axis}
\end{tikzpicture}

%% file: Journal/sections/framework_general.tex
\section{Byzantine-Resilient Gradient Coding}
\label{sec:framework}

\input{Journal/sections/notation}

Our solution is inspired by the example in \cref{sec:example}. Carrying over that idea of an interactive game on the problem of Byzantine errors in distributed gradient descent, we next define the following framework of gradient coding schemes resilient against an adversary controlling $\nmalicious$ malicious workers.

\begin{definition}[Byzantine-resilient gradient coding scheme]
    A Byzantine-resilient gradient coding scheme tolerating $\nmalicious$ malicious workers, referred to as \BGC, is a tuple $\left( \allocmat, \encfunset, \decfun, \proto \right)$ where
\begin{itemize}
    \item $\allocmat \in \{0, 1\}^{\ngrad \times \nworker}$ is a \textbf{data assignment matrix} in which $\allocmat[{\gind, \wind}]$ is equal to $1$ if the $\gind$-th data sample is given to the $\wind$-th worker and $0$ otherwise,
    \item $\encfunset \defeq \left(
        \encfun[\wind,\encind] \colon \galpha^{\graddim \times \ngrad} \to \{0, 1\}^{\respdim{\encind}{\wind}}
        \mid \wind\in \range{\nworker}, \encind \in \range{\nencfun}
    \right)$ is the list of $\nworker\nencfun$ \textbf{encoding functions} used by the workers such that $\encfun[\wind,1]$ corresponds to a gradient code dictated by $\allocmat$ and $\encfun[\wind,\encind]$ depends only on the gradients assigned to $\worker$,
    \item $\proto = (\proto_1,\proto_2)$ is a \textbf{multi-round protocol} in which $\proto_1$ selects the indices of the encoding functions to be used by the workers based on all previous responses at the start of each iteration, and $\proto_2$ selects gradients to be locally computed at the \master based on all previous responses at the end of each iteration,
    \item and $\decfun$ is a \textbf{decoding function} used by the \master after running the protocol $\proto$ to always output the correct full gradient if the number of malicious workers is at most $\nmalicious$.
\end{itemize}
\end{definition}

Each worker initially ($\roundind=0$)
sends a vector $\iresp_{0,\wind} \defeq \encfun[\wind,1]\left( \pgrad[1],\dots,\pgrad[\ngrad] \right) \in \galpha^{\graddim}$ that is a codeword symbol of a gradient code~\cite{tandonGradientCoding2017}. The protocol $\proto$ then runs for $\nround \in \N$ rounds.

At start of each round $\roundind \in \range{\nround}$, the \master uses $\proto_1$ to select an encoding function $\encfun[\wind,\encind_{\roundind,\wind}]$ for each worker $\worker$ and communicates its index $\encind_{\roundind,\wind}$ to the respective worker.
Each worker $\worker$ then computes a response\footnote{Strictly speaking, the encoding function outputs bits. Values from $\galpha$ are assigned an arbitrary but fixed order and the binary representation of the 
corresponding index is transmitted.}
\begin{equation*}
\iresp_{\roundind,\wind} \defeq \encfun[\wind,\encind_{\wind,\roundind}]\left( \pgrad[1],\dots,\pgrad[\ngrad] \right)
\in \galpha^{\respdim{\encind_{\wind,\roundind}}{\wind}}
\end{equation*}
and sends a vector $\noisyiresp_{\roundind,\wind} \in \galpha^{\encdim{\wind}{\encind_{\wind,\roundind}}}$ to the \master.
For honest workers $\noisyiresp_{\roundind,\wind} = \iresp_{\roundind,\wind}$, while for malicious workers, $\noisyiresp_{\roundind,\wind}$ may be chosen arbitrarily.

At the end of each round, the \master uses $\proto_2$ to choose a set of partial gradients to compute locally. We denote the list of indices of the locally computed partial gradients in round $\roundind$ by $\lgindset_\roundind$ and the list of corresponding partial gradient values by $\locgradset_\roundind \defeq (\pgrad[\gind] \mid \gind \in \lgindset_\roundind )$.
Analogously, we define $\irespset_\roundind \defeq (\iresp_{\roundind, \wind} \mid \forall \wind \in [\nworker])$ and
$\recresset_\roundind \defeq (\noisyiresp_{\roundind, \wind} \mid \forall \wind \in [\nworker])$.

The protocol $\proto_1$ selects the indices of the encoding functions to be used in the current round $t$ based on the received results and locally computed gradients from previous rounds.
After receiving the results in the current round, the \master uses $\proto_2$ to select the gradients to compute locally in this round, i.e.,
\vspace{-0.5em}
\begin{align}
    \encind_{1,\roundind}, \dots, \encind_{\nworker, \roundind} &= \proto_1\left(
        \recresset^{(\roundind-1)},
        \lgindset^{(\roundind-1)},
        \locgradset^{(\roundind-1)}
    \right), \label{eq:encind}\\
    \lgindset_\roundind &= \proto_2\left(
        \recresset^{(\roundind)},
        \lgindset^{(\roundind-1)},
        \locgradset^{(\roundind-1)}
    \right) \label{eq:indset}.
\end{align}
After round $\nround$, the main node computes an estimate $\gestim$ of $\tgrad$ using the decoding function 
    $\gestim = \decfun\left(\recresset^{(\nround)}, \lgindset^{(\nround)}, \locgradset^{(\nround)} \right).$
For convenience and later reference, we summarize the symbol definitions in \cref{tab:notation}.

We study settings in which the number of workers \(\nworker\) is a multiple of \(\nmalicious+\nhonest\), i.e., \(\nworker = \ngroup (\nmalicious + \nhonest)\) for some integers \(\ngroup,\nhonest \geq 1\) and only consider balanced data assignments, i.e., every worker computes the same number of gradients.
We focus on the particular case of a fractional repetition data assignment~\cite{tandonGradientCoding2017}. That is, the \master partitions the workers into \(\ngroup\) groups of size \(\frac{\nworker}{\ngroup}\) each and assigns the same data samples to all workers within a group.
The data assignment matrix is constructed as 
\ifarxiv \else 

\vspace{-1em}
\fi
\begin{equation}
\label{eq:fractional_repetition}
\allocmat = \begin{bmatrix}
        \boldsymbol{1}_{\frac{\ngrad}{\ngroup} \times \frac{\nworker}{\ngroup}} & \boldsymbol{0}_{\frac{\ngrad}{\ngroup} \times \frac{\nworker}{\ngroup}} & \dots & \boldsymbol{0}_{\frac{\ngrad}{\ngroup} \times \frac{\nworker}{\ngroup}} \\
        \boldsymbol{0}_{\frac{\ngrad}{\ngroup} \times \frac{\nworker}{\ngroup}} & \boldsymbol{1}_{\frac{\ngrad}{\ngroup} \times \frac{\nworker}{\ngroup}} &   \dots & \boldsymbol{0}_{\frac{\ngrad}{\ngroup} \times \frac{\nworker}{\ngroup}} \\
        \vdots & \vdots & \ddots & \vdots \\
        \boldsymbol{0}_{\frac{\ngrad}{\ngroup} \times \frac{\nworker}{\ngroup}} & \boldsymbol{0}_{\frac{\ngrad}{\ngroup} \times \frac{\nworker}{\ngroup}} & \dots & \boldsymbol{1}_{\frac{\ngrad}{\ngroup} \times \frac{\nworker}{\ngroup}}
\end{bmatrix}.
\end{equation}
Since we focus on this particular data assignment, the initial worker responses are given by the sum of all computed gradients
$\iresp_{0,\wind} =
\sum_{
\substack{\gind \in \range{\ngrad}, \allocmat[\gind,\wind]=1}
} \pgrad[\gind]$
, which form a valid gradient code.
\\
An \BGC scheme is evaluated by the maximum number of rounds $\nround$, the maximum number of local computations required by $\proto$, its replication factor, and communication overhead, all of which we formally define next.

\begin{definition}%
The total number of partial gradients obtained by \textbf{local computations} at the \master is defined as \begin{equation*}
    \localcomp \defeq \card{\lgindset^{(\nround)}}.
\end{equation*}

The \textbf{replication factor} of an \BGC scheme is the average number of workers to which each sample is assigned, i.e.,
\begin{equation*}
    \replfact \defeq \frac{\sum_{\gind \in \range{\ngrad}, \wind \in \range{\nworker}} \allocmat[{\gind, \wind}]}{\ngrad}.
\end{equation*}

The \textbf{communication overhead} is the maximum number of bits transmitted from the workers to the \master during $\proto$, i.e.,
\begin{equation*}
    \commoh \defeq \sum_{\roundind \in \range{\nround}, \wind \in \range{\nworker}} \respdim{\encind_{\wind,\roundind}}{\wind}.
\end{equation*}
\end{definition}

We say that a tuple $(\allocmat, \encfunset, \decfun, \proto)$ is a \BGC scheme with parameters (\nround,\localcomp,\replfact,\commoh) if in the presence of at most $\nmalicious$ malicious workers, the scheme always outputs $\gestim = \tgrad$ and requires at most $\nround$ communication rounds, at most $\localcomp$ local computations, a replication factor $\replfact$, and a communication overhead of at most $\commoh$.

%% file: Journal/sections/notation.tex
{
\begin{table*}[t]
\caption{Overview of symbol definitions}
\label{tab:notation}
\begin{tabularx}{\linewidth}{lX lX lX}
\cmidrule(lr){1-2}
\cmidrule(lr){3-4}
\cmidrule(lr){5-6}
$\nworker$ & number of workers
    & $\encfun[\wind,\encind]$ & $\encind$-th encoding function at worker $\wind$
        & $\allocmat$ & data allocation matrix of size $\ngrad \times \nworker$
\\
$\nmalicious$ & number of malicious workers
    & $\decfun$ & decoding function used by the \master
        & $\recresset_{\roundind}$ & list of responses in round $\roundind$
\\
$\nhonest$ & number of honest workers per group %
    & $\proto$ & interactive protocol
        & $\irespset_{\roundind}$ & list of honest responses in round $\roundind$
\\
$\ngroup$ & number of fractional repetition groups %
    & $\replfact$ & replication factor per partial gradient
        & $\lgindset_{\roundind}$ & list of indices for gradients computed at the \master in round $\roundind$
\\
$\ngrad$ & number of partial gradients
    & $\commoh$ & communication overhead of protocol $\proto$
        & $\locgradset_{\roundind}$ & list of values for gradients computed at the \master in round $\roundind$
\\
$\tgrad$ & correct total gradient %
    & $\localcomp$ & number of gradient computations at the \master
        & $\disagreegradset$ & set of indices for corrupted gradients in the symmetrization attack in \cref{sec:attack} 
\\
$\gestim$ & decoded total gradient %
    & $\nround$ & number of rounds in the protocol $\proto$
        & $\encfunset$ & list of encoding functions
\\
$\pgrad$ & correct partial gradient for sample $\sample$
    & $\roundind$ & round index for the protocol $\proto$
        & $\galpha$ & message alphabet
\\
$\wpgrad{\gind}{\wind}$ & claimed value for $\pgrad$ from worker $\wind$ %
    & $\gditer$ & index of the gradient descent iteration
        & &
\\
\cmidrule(lr){1-2}
\cmidrule(lr){3-4}
\cmidrule(lr){5-6}
\end{tabularx}
\end{table*}}

%% file: Journal/sections/fundamentals.tex
\section{Fundamental Limits of \BGC Schemes}
\label{sec:fundamental_limits}

This section contains detailed proofs of the fundamental limits on the number of local computations at the \master (\cref{thm:converse_C}) and the communication overhead (\cref{thm:converse_commoh}). The underlying methods and ideas are presented. The lower bounds are partly based the adversary adopting a specific strategy, which is why we first present this strategy in the following.

\subsection{Symmetrization Attack}
\label{sec:attack}
We base parts of our proofs on the concept of a symmetrization attack. The core idea is for the adversary to choose errors such that the \master cannot distinguish between different cases.
As the adversary controls $\nmalicious$ workers, it can choose a (potentially corrupted) value for each malicious worker and each partial gradient.
We denote worker $\worker$'s \emph{claimed} partial gradient results for $\pgrad$ as $\wpgrad{\gind}{\wind}\in\galpha^\graddim$ for all $\wind \in \range{\nworker}$ and $\gind\in\range{\ngrad}$.
For honest workers we say $\wpgrad{\gind}{\wind} = \pgrad[\gind]$.
Each worker $\worker$ computes their responses consistently based on those values, i.e.,
\ifarxiv \else 

\vspace{-1em}
\fi
\begin{align*}
    \noisyiresp_{\roundind, \wind} &=
    \encfun[\wind,\encind_{\wind,\roundind}]\left(
        \wpgrad{1}{\wind},\dots,\wpgrad{\ngrad}{\wind}
    \right).
\end{align*}
For clarity of exposition, we first lay out how the adversary chooses the claimed gradient values for $\ngroup=1$ groups of size $\nmalicious+1$, i.e., for $\nhonest=1$, before we generalize to arbitrary $\ngroup, \nhonest \in \N$.

For $\nhonest=1$ and $\ngroup=1$, the adversary draws a set $\disagreegradset \subseteq \range{\ngrad}$ of size $|\disagreegradset|=\lfloor \frac{\nmalicious}{\nhonest} \rfloor = \nmalicious$ uniformly at random and assigns each malicious worker $\worker$ a unique gradient index from $\disagreegradset$. The adversary will introduce errors only for gradients $\pgrad[\tilde{\gind}]$, $\tilde{\gind} \in \disagreegradset$ and only for the one malicious worker that got assigned gradient index $\tilde{\gind}$.
With probability $\frac{1}{2}$, it picks a single gradient index from $\disagreegradset$ uniformly at random and sets all malicious workers' values to the same erroneous random partial gradient value.
The resulting claimed gradients take on the form as depicted in \cref{tab:key_error_pattern}.
For partial gradients with index in $\disagreegradset$, there are two competing values $\pgrad^{\prime}$ and $\pgrad^{\prime\prime}$ whereas for all other gradient indices, the claimed values by all workers agree.

\begin{table}[h]
    \centering
    \caption{claimed partial gradients for symmetrization attack}
    \ifarxiv
        \begin{tabularx}{0.88\columnwidth}{c|ccccccc}
            & $\wpgrad{1}{\wind}$ & $\wpgrad{2}{\wind}$ 
            & $\dots$  & $\wpgrad{\nmalicious}{\wind}$ & $\wpgrad{\nmalicious+1}{\wind}$ & $\dots$ & $\wpgrad{\ngrad}{\wind}$ \\ 
            \toprule
            $\worker[1]$ & $\pgrad[1]^{\prime\prime}$ & $\pgrad[2]^\prime$ & $\dots$  & $\pgrad[\nmalicious]^\prime$ & $\pgrad[\nmalicious+1]^\prime$ & $\dots$ & $\pgrad[\ngrad]^\prime$ \\ 
            $\worker[2]$ & $\pgrad[1]^\prime$ & $\pgrad[2]^{\prime\prime}$ & $\dots$  & $\pgrad[\nmalicious]^\prime$ & $\pgrad[\nmalicious+1]^\prime$ & $\dots$ & $\pgrad[\ngrad]^\prime$ \\ 
            $\vdots$ & $\vdots$ & $\vdots$ & $\ddots$ & $\vdots$ & $\vdots$ & $\dots$ & $\vdots$ \\
            $\worker[\nmalicious]$ & $\pgrad[1]^\prime$ & $\pgrad[2]^\prime$ & $\dots$ & $\pgrad[\nmalicious]^{\prime\prime}$ & $\pgrad[\nmalicious+1]^\prime$ & $\dots$ & $\pgrad[\ngrad]^\prime$ \\ 
           $\worker[\nmalicious+1]$ & $\pgrad[1]^\prime$ & $\pgrad[2]^\prime$ & $\dots$  & $\pgrad[\nmalicious]^\prime$ & $\pgrad[\nmalicious + 1]^\prime$ & $\dots$ & $\pgrad[\ngrad]^\prime$ \\ 
        \end{tabularx}
    \else
        \begin{tabularx}{0.5\columnwidth}{c|ccccccc}
            & $\wpgrad{1}{\wind}$ & $\wpgrad{2}{\wind}$ 
            & $\dots$  & $\wpgrad{\nmalicious}{\wind}$ & $\wpgrad{\nmalicious+1}{\wind}$ & $\dots$ & $\wpgrad{\ngrad}{\wind}$ \\ 
            \toprule
            $\worker[1]$ & $\pgrad[1]^{\prime\prime}$ & $\pgrad[2]^\prime$ & $\dots$  & $\pgrad[\nmalicious]^\prime$ & $\pgrad[\nmalicious+1]^\prime$ & $\dots$ & $\pgrad[\ngrad]^\prime$ \\ 
            $\worker[2]$ & $\pgrad[1]^\prime$ & $\pgrad[2]^{\prime\prime}$ & $\dots$  & $\pgrad[\nmalicious]^\prime$ & $\pgrad[\nmalicious+1]^\prime$ & $\dots$ & $\pgrad[\ngrad]^\prime$ \\ 
            $\vdots$ & $\vdots$ & $\vdots$ & $\ddots$ & $\vdots$ & $\vdots$ & $\dots$ & $\vdots$ \\
            $\worker[\nmalicious]$ & $\pgrad[1]^\prime$ & $\pgrad[2]^\prime$ & $\dots$ & $\pgrad[\nmalicious]^{\prime\prime}$ & $\pgrad[\nmalicious+1]^\prime$ & $\dots$ & $\pgrad[\ngrad]^\prime$ \\ 
           $\worker[\nmalicious+1]$ & $\pgrad[1]^\prime$ & $\pgrad[2]^\prime$ & $\dots$  & $\pgrad[\nmalicious]^\prime$ & $\pgrad[\nmalicious + 1]^\prime$ & $\dots$ & $\pgrad[\ngrad]^\prime$ \\ 
        \end{tabularx}
    \fi
    \label{tab:key_error_pattern}
\end{table}

For $\nhonest > 1$, the adversary randomly partitions the malicious workers into
$\lfloor \nmalicious / \nhonest \rfloor -1$ sets of size $\nhonest$ and (depending on divisibility) one group of size $\nmalicious \mod \nhonest$.
Each of the sets of size $\nhonest$ behaves like one malicious worker in the case $\nhonest=1$.
The workers in the remaining set of less than $\nhonest$ workers (if non-empty) pick their claimed gradients randomly either like a random other set of workers or like the honest workers.
The resulting claimed gradients take on the form as depicted in \cref{app:large_table}.

For $\ngroup > 1$, the adversary chooses the claimed gradients in the first group according to the attack strategy for $\ngroup^\prime=1$ and $\ngrad^\prime=\ngrad / \ngroup$. In all other groups, the claimed gradients equal their true values.

\subsection{Limits on Computation}
\label{subsec:limits_computation}

\begin{IEEEproof}[Proof of \cref{thm:converse_C}]
We demonstrate that any tuple $(\allocmat, \encfunset, \decfun, \proto)$ with fractional repetition data allocation which uses $\localcomp < \compbound$ local computations cannot be an \BGC scheme. Specifically, we show that the malicious workers can always perform a symmetrization attack, preventing the main node from deterministically computing the full gradient.
Suppose that all malicious workers are contained in the same group. Since, for fractional repetition, the groups are independent of each other, we only focus on a single group. Let the malicious workers behave as explained in \cref{sec:attack}.

We abstract the communication and $\proto_1$ by assuming that the values of all partial gradients $\wpgrad{\gind}{\wind}$ claimed by the workers are available at the \master. Note that, regardless of the particular communication and $\proto_1$, the \master cannot gain any additional information from the workers' responses.
We now show that for $\localcomp < \compbound$ there exists no choice of a decoding function $\decfun$ and $\proto_2$ for which the \master deterministically outputs the true full gradient. To that end, for any possible $\proto_2$, we give two cases for which the inputs to the decoding function are identical, but the true full gradients differ.

The claimed partial gradients take on values as in \cref{tab:key_error_pattern} for $\nhonest=1$ and as in \cref{app:large_table} for $\nhonest>1$. There are $\compbound$ partial gradients, w.l.o.g. say $\pgrad[1],\ldots,\pgrad[\compbound]$, such that for each of these partial gradients, there are $\nhonest$ workers that claim a value $\wpgrad{\gind}{\wind} = \pgrad[\gind]^{\prime\prime}$ that is different from the value $\pgrad[i]^\prime$ claimed by all other workers, where $\gind=1,\ldots,\compbound$. That is, as explained in \cref{sec:attack}, the malicious workers choose a strategy by splitting into subgroups of size $\nhonest$ and aligning their claimed partial gradient values within each subgroup.
For any list $\lgindset^{(\nround)}$ of locally computed gradients of size $|\lgindset^{(\nround)}| < \compbound$ (output by any $\proto_2$) there exists an index $\widetilde{\gind} \in [\compbound]$ such that $\widetilde{\gind} \notin \lgindset^{(\nround)}$.
Consider the following two cases:\begin{enumerate}[label=Case \arabic{*}:, wide=1.5\parindent, leftmargin=*]
    \item $\pgrad[\gind] = \pgrad[\gind]^\prime \;\forall \gind \in [\frac{\ngrad}{\ngroup}]$ and 
    \item $\pgrad[\gind] = \pgrad[\gind]^\prime \;\forall \gind \in [\frac{\ngrad}{\ngroup}] \setminus \{\widetilde{\gind}\}$ and $\pgrad[\widetilde{\gind}] = \pgrad[\widetilde{\gind}]^{\prime\prime}$.
\end{enumerate}
It is easy to see that both cases occur with non-zero probability according to the attack strategy in \cref{sec:attack}. In both cases, the inputs to $\decfun$ only depend on the list $\lgindset^{(\nround)}$, the claimed values $\wpgrad{\gind}{\wind},\;\wind \in [\nmalicious+\nhonest],\;\gind\in[\frac{\ngrad}{\ngroup}]$ and the locally computed values $\pgrad[\gind],\; \gind \in \lgindset^{(\nround)}$, all of which take on identical values in both cases; that is, the two cases are not distinguishable at the decoder.
The value of the full gradient $\tgrad$, however, is $\sum_{\gind \in [\ngrad]} \pgrad[\gind]^\prime$ in Case 1 and $\pgrad[\widetilde{\gind}]^{\prime\prime} + \sum_{\gind \in [\ngrad] \setminus \{\widetilde{\gind}\}} \pgrad[\gind]^\prime$ in Case 2. Hence, no decoding function can deterministically produce the correct full gradient for $\localcomp < \compbound$.
\end{IEEEproof}

\begin{IEEEproof}[Extension to stragglers (\cref{cor:stagglers})]
  We extend our result to the straggler case by the following observation. Suppose that $\nstraggle$ workers are stragglers, i.e., they are not malicious but don't report anything to the \master, where $0 \leq \nstraggle < \nhonest$. Consider the case in which all stragglers and malicious workers are contained in the same fractional repetition group. The computations in all other fractional repetition groups are independent. Thus, the minimum number of local computations is equal to the minimum number of local computations in a setting with $\nmalicious$ malicious workers and $\nhonest-\nstraggle$ honest workers per group, which implies that $\localcomp \geq \left\lfloor \frac{\nmalicious}{\nhonest-\nstraggle} \right\rfloor$.
\end{IEEEproof}

\subsection{Limits on Communication}
\label{subsec:limits_communication}
\input{Journal/proofs/lower_bound_commoh}

%% file: Journal/proofs/lower_bound_commoh.tex
For the proof of \cref{thm:converse_commoh} we utilize the following direct consequence of the proof technique used to prove \cref{thm:converse_C}.

\begin{corollary}[Computation of Disagreement Gradients]\label{cor:disagreement}
    For any \BGC scheme with parameters (\nround,\localcomp,\replfact,\commoh) that has a fractional repetition data assignment with $\replfact = \nmalicious + \nhonest$,
    if the adversary employs a symmetrization attack according to \cref{sec:attack}, then the list of locally computed gradients must contain the list of all gradients on which the workers disagree, i.e.,
  $\forall \widetilde{\gind} \in \disagreegradset:  \widetilde{\gind} \in \lgindset^{(\nround)}$.%
    \label{cor:comp_disag_grads}
\end{corollary}
\begin{IEEEproof}
    The proof follows the same steps as for \cref{thm:converse_C}. Note that if there exists an index $\widetilde{\gind} \in \disagreegradset: \widetilde{\gind} \notin \lgindset^{(\nround)}$, there are two cases that cannot be distinguished based on the information available at the \master.
\end{IEEEproof}

\begin{IEEEproof}[Proof of \cref{thm:converse_commoh}]
    We consider a single group consisting of $\nworker = \nmalicious + \nhonest$ workers. Since the datasets per group, as well as the sets of workers, are disjoint, the communication necessary for $\ngroup$ groups of size $\ngrad/\ngroup$ is at least as big as for $\ngroup^\prime=1$ group of size $\ngrad/\ngroup$.
    Further, we assume the behavior of the malicious workers as in \cref{sec:attack}.
    According to \cref{cor:comp_disag_grads}, for $\localcomp = \lfloor \frac{\nmalicious}{\nhonest} \rfloor$, we require every item of $\disagreegradset$ to be in $\lgindset^{(\nround)}$, while $\localcomp = |\lgindset^{(\nround)}| = |\disagreegradset| = \compbound$.
    In other words, the \master must exactly compute all the gradients in $\disagreegradset$ locally. If there is a non-zero probability that a different gradient is computed locally, the scheme cannot be a valid \BGC scheme.
    Overall, $\disagreegradset$ must be uniquely determined by the \master's available information at the end of the protocol, leading to
        $\En(\disagreegradset \given \wtmdata, \locgradset^{(\nround)}) = 0$,
    where $\wtmdata$ denotes the list of random variables corresponding to all data transmitted from all workers to the \master and $\locgradset^{(\nround)}$ denotes the list of the random variables corresponding to the values of the locally computed gradients.
    Using this we have
\ifarxiv \else 

\vspace{-1em}
\fi
    \begin{align}
        \En(\wtmdata \given \locgradset^{(\nround)}) 
                    \label{eq:defmi}
                    &\geq \I(\wtmdata; \disagreegradset \given \locgradset^{(\nround)}) \\ 
                    \notag
                    &= \En(\disagreegradset \given \locgradset^{(\nround)}) - \En(\disagreegradset \given \locgradset^{(\nround)}, \wtmdata) \\
                    \notag
                    &= \En(\disagreegradset \given \locgradset^{(\nround)}) \\
                    \label{eq:indep_disagreegradset}
                    &= \En(\disagreegradset) \\
                    &= \log_2 \binom{\ngrad / \ngroup}{\lfloor \nmalicious / \nhonest \rfloor},\label{eq:uniform}
    \end{align}
    where \eqref{eq:defmi} follows from the definition of mutual information, 
    \eqref{eq:indep_disagreegradset} follows since $\disagreegradset$ is independent from $\locgradset^{(\nround)}$ and \eqref{eq:uniform} holds since $\disagreegradset$ is a uniform selection of $\localcomp = \lfloor \nmalicious / \nhonest \rfloor$ indices out of $[\ngrad / \ngroup]$.
    To transmit the information in $\wtmdata$ with zero error to the \master, the workers need to send at least $\log_2 \binom{\ngrad / \ngroup}{\lfloor \nmalicious / \nhonest \rfloor}$ bits.

\end{IEEEproof}
\begin{IEEEproof}[Extension to stragglers (\cref{cor:stagglers})]
  The same line of arguments applies to the straggler case. As established in \cref{subsec:limits_computation}, in the presence of $\nstraggle$ stragglers, it is necessary to find a larger set $\disagreegradset$ of disagreeing gradient indices; namely, $\localcomp = \card{\lgindset^{(\nround)}} = \card{\disagreegradset} = \left\lfloor \frac{\nmalicious}{\nhonest-\nstraggle} \right\rfloor$. Applying the same ideas gives
  \ifarxiv \else

  \vspace{-1em}
  \fi
  \begin{equation*}
    \En(\wtmdata \given \locgradset^{(\nround)})
    \geq
    \log_2 \binom{\ngrad / \ngroup}{\lfloor \nmalicious / (\nhonest-\nstraggle) \rfloor}.
  \end{equation*}
\end{IEEEproof}

%% file: Journal/sections/scheme.tex
\section{Construction and Analysis of an \BGC Scheme}
\label{sec:scheme}

We construct an \BGC scheme with parameters (\nround, \localcomp, \replfact, \commoh) that has a replication factor \mbox{$\replfact={\nmalicious+\nhonest}$} for \mbox{$\nhonest \geq 1$} and achieves the optimal local computation load \mbox{$\localcomp \leq \compbound$} at the \master. The protocol $\proto$ runs for
\mbox{$\nround
    \leq
    \left( \nmalicious - \bar{\localcomp} (\nhonest-1) \right)
    \left( 2\left\lceil \log_2\left(\frac{\ngrad}{\ngroup}\right) \right\rceil + 1 \right)$
}
rounds and achieves an asymptotic communication overhead
\mbox{$\commoh \leq
  \left( \nmalicious-\bar{\localcomp}(\nhonest-1) \right)
  \left( 1+ \left\lceil \log_2{\card{\galpha}} \right\rceil \right)
\left\lceil \log_2\left( \frac{\ngrad}{\ngroup}\right) \right\rceil$}
. Our scheme uses a fractional repetition data assignment with $\ngroup$ groups of size $\nmalicious+\nhonest$. Informally, in each group, the \master runs an elimination tournament consisting of matches (similar to the one explained in~\cref{sec:example}) between pairs of workers that return contradicting responses.
For clarity of exposition, we explain the idea of our scheme for the special case of $\nhonest=1$ first and then describe the general case.
In the following, the elimination tournament is explained for a single fractional repetition group. The \master repeats the same procedure for each fractional repetition group subsequently.

\subsection{Special case for $\nhonest =1$}
\input{Journal/sections/special_case}

\subsection{General case for any $\nhonest \geq 1$}
\label{subsec:general_construction}
\input{Journal/sections/general_case}

\subsection{Properties of the framework}
\input{Journal/proofs/achievability_scheme}

\section{Discussion}
\label{sec:discussion}

According to \cref{thm:converse_C}, the interactive protocol of our scheme achieves the lowest possible number of locally computed partial gradients for the fractional repetition data assignment and $\replfact = \nmalicious+\nhonest$. Note that for $\nhonest \geq \nmalicious+1$, i.e., $\replfact \geq 2\nmalicious+1$, no local computation is necessary. In fact, since there is only one set of consistent workers $\workerset_\ell$ that has $\card{\workerset_\ell} \geq \nhonest$ per fractional repetition group, the scheme immediately identifies $\workerset_\ell$ as the honest set and terminates without any additional computation or communication. Thus, as shown in~\cite{chenDRACOByzantineresilientDistributed2018}, this is optimal. Although we consider $1 \leq \nhonest \leq \nmalicious+1$ in \cref{thm:scheme} for technical reasons, the scheme works for any $\nhonest \geq 1$. We remark that since $\ngrad \gg \nmalicious$ in state-of-the-art machine learning deployments, the local computations cause only a relatively small load at the \master.

\begin{figure}[h]
  \centering
  \ifarxiv
  \resizebox{0.85\linewidth}{!}{ \input{tikz/comm_vs_localcomp_updated2.tex} }
  \else
  \resizebox{0.5\linewidth}{!}{ \input{tikz/comm_vs_localcomp_updated2.tex} }
  \fi
  \caption{Tradeoff between total worker to main-node communication and the number of local computations for our scheme. The parameters are $\nmalicious = 10$, $1 \leq \nhonest \leq 11$, $\ngroup=1$, $\ngrad = \num{1e4}$, $\graddim=\num{1e6}$, $|\galpha|=2^{16}$. When considering total communication, the cost of the protocol is outweighed by the cost of transmitting the gradient values. The gap to our bound, given in \cref{thm:converse_commoh}, is less than \SI{5}{\kilo\byte}.
  For $\localcomp=0$ local computations, our scheme is equivalent to DRACO~\cite{chenDRACOByzantineresilientDistributed2018}.
  By requiring fewer workers, our scheme with $\nhonest=1$ reduces communication by \SI{48}{\percent} at the expense of at most $\localcomp=10$ local gradient computations at the \master. For comparison, each worker node performs $\ngrad=\num{1e4}$ gradient computations.
  }
  \label{fig:comm_vs_localcomp}
\end{figure}

As depicted in \cref{fig:comm_vs_localcomp}, for realistic parameter ranges, the local computations drastically reduce the required communication. The communication overhead $\commoh$ of the protocol is outweighed by the initial transmission of $\iresp_{0, \wind}$.
Additionally, we remark that although the communication complexity of our scheme is quadratic in $\nmalicious$, this value is not very large in practice.

Note that the achievable bound on $\commoh$ in \cref{thm:scheme} is off from the converse bound in \cref{thm:converse_commoh} by a constant factor asymptotically, see \cref{fig:achievability_vs_converse_datacolumns}. The reason is that we use a rather simple and conservative lower bound on the amount of information that is required to be transmitted by the workers in \cref{thm:converse_commoh}. 
The main limitations of the converse bound are as follows. First, we assume very specific behavior for the malicious workers, especially we assume they are cooperative in transmitting the indices of the gradients they corrupted. Second, the bound neglects the fact that the locally computed partial gradient values need to be compared to a worker's claimed partial gradient values at the main node. The interplay between these two pieces of information, together with workers' malicious behavior, is not trivial.

\input{Journal/sections/achievability_vs_converse}

Our gains in terms of replication over previous works are mainly driven by two relaxations: we allow local computations at the \master and interactive communication, which, in particular, involves feedback from the \master to the workers. The amount of data that is conveyed through the feedback in our scheme is in the order of the amount of data that the challenging user transmits. In turn, the voting users transmit only messages of constant size once per elimination round. This fraction of the communication overhead does not grow with $\ngrad$ and is negligible for large datasets (see \cref{cor:achievable_commoh}). The uplink and downlink, therefore, have a symmetric communication load. In particular, the downlink communication consists of two components. At every stage of a match, the \master first needs to communicate the descent direction in the match tree (proceed to the left child or right child), which is a binary message to the dedicated challenging worker. Second, after having received a proposed value for the node label from the challenging worker, the \master needs to send this value to the voting worker. While this message is a unicast most of the time, for $\nhonest > 1$ at the last stage of a match, the proposed node label is sent to all workers in the competing worker groups by a multi-cast. Having at most $\nmalicious-\bar{\localcomp}(\nhonest-1)$ matches, the total number of bits communicated in the downlink is at most $\left( \nmalicious-\bar{\localcomp}(\nhonest-1) \right) \left( 1 + \left\lceil \log_2\card{\galpha} \right\rceil \right)$.
It is possible to reduce the data that is sent over the feedback channel by increasing the uplink communication. For example, instead of voting for a proposed gradient, each user can transmit their own proposal, and the \master can compare the proposals locally. The exploration of the fundamental trade-off between uplink and downlink communication is an open problem for future research.

%% file: Journal/sections/special_case.tex
We formalize the idea that was presented in \cref{sec:example}.
The \master runs an elimination tournament consisting of a series of matches between two workers.
During a match, each worker constructs a binary tree based on their computed partial gradients, which we refer to as the match tree.
The root of the tree is labeled by the sum of all partial gradients computed at that worker ($\noisyiresp_{0,\wind}$).
The child nodes are constructed based on the partial gradients that are contained in the parent node.
Each node has two children: the first one is labeled by the sum of the first half of the parent node's partial gradients; the second one is labeled by the sum of the second half of the parent node's partial gradients.
Thus, the labeling is done such that the sum of the labels of any two siblings gives the label of their parent node.

Proceeding in this way recursively, each worker ends up with the leaves of the tree being labeled by individual partial gradients.
For example, when $\ngroup = 1$ and $\ngrad=4$, the tree is depicted in \cref{fig:binary_add_tree}.
During a match, the \master compares the labels for particular nodes in this tree from the two competing workers.
If the root labels of two match trees differ, then there must be a child node for which the corresponding label differs between those trees. By induction, it is clear that there has to be a path from the root to a leaf, such that the corresponding labels of all involved nodes differ between the two match trees. Applying this observation to the example in \cref{fig:binary_add_tree}, if $\wpgrad{1}{\wind}+\wpgrad{2}{\wind}+\wpgrad{3}{\wind}+\wpgrad{4}{\wind}$ is different for two workers, then $\wpgrad{1}{\wind}+\wpgrad{2}{\wind}$ or $\wpgrad{3}{\wind}+\wpgrad{4}{\wind}$ must also differ (or both). In the latter case, we end up with $\wpgrad{3}{\wind}$ or $\wpgrad{4}{\wind}$ being different between the workers.
\begin{figure}[t]
    \centering
    \ifarxiv
        \resizebox{0.9\linewidth}{!}{ \input{tikz/matchtree.tex} }
    \else
        \resizebox{0.5\linewidth}{!}{ \input{tikz/matchtree.tex} }
    \fi
  \vspace{-1em}
    \caption{Example of a match tree for $\worker$ and parameters $\ngroup=1,\ngrad=4$.}
    \label{fig:binary_add_tree}
    \vspace{-0.5cm}
\end{figure}

The match starts at the root. In this case, each worker $\worker$ would have already sent the node label in $\iresp_{0,\wind}$ to the \master, i.e.,
\ifarxiv \else 

\vspace{-1em}
\fi
\begin{equation*}
    \iresp_{0,\wind} = \encfun[\wind,1]\left( \pgrad[1],\dots,\pgrad[\ngrad] \right) = \sum_{\substack{ \gind \in \range{\ngrad} \\ \allocmat[\gind, \wind] = 1}} \pgrad[\gind].
\end{equation*}
Note that, without errors, all workers' messages agree within a group.
In case of discrepancies between the responses $\noisyiresp_{0,\wind}$ of the workers within a group, the \master selects a pair of disagreeing workers $\worker[\wind_1],\worker[\wind_2]$ and further descends in the tree as follows.

For every node in the tree, the \master requests the left child's label from one of the workers (\emph{challenging worker}) and asks the second worker to either support the claimed value or reject it by sending a single bit (\emph{voting worker}). The responses are encoded in the messages $\iresp_{\roundind,\indone}$ and $\iresp_{\roundind,\indtwo}$ respectively.
Based on the current node's label and the left child's label, the \master can infer the right child's label also: if the competing workers agree on the left child's label, they must disagree on the right child's label.
The \master then moves on to a child whose label the workers disagree on.
This procedure is repeated until a leaf is reached.
Note that, even if the workers send inconsistent responses each round, this procedure is guaranteed to reach a leaf node for which the values of the individual partial gradient differs between for the two workers.

It is possible to reduce the communication load by picking a coordinate $\compind$ in which the workers' initial responses disagree, i.e., $\comp{ \iresp_{0,\indone} } \neq \comp{ \iresp_{0,\indtwo} }$. That is, the procedure operates only on the $\compind$-th coordinate of the node labels. By this method, it is still guaranteed that the \master can identify a partial gradient for which the competing workers disagree.

Having identified disagreeing values $\wpgrad{\gind}{\wind_1}$ and $\wpgrad{\gind}{\wind_2}$ of a partial gradient $\pgrad[\gind]$, the \master computes the correct value of this partial gradient locally. It then marks one of the workers as malicious: either the challenging worker reported a wrong value, or the voting worker rejected a correct value.%
\footnote{If the challenging worker is exposed as a liar, having rejected a wrong value does not mean that the non-eliminated worker is honest. The scheme can be modified to also query the partial gradient value from the voting worker, which allows the \master to eliminate both workers at the same time in some cases. Since we design our scheme for worst-case guarantees, however, we achieve a smaller communication overhead from workers to \master by our strategy.}
The algorithm ends up with disagreeing leaf labels by design. Each match is guaranteed to eliminate at least one malicious worker and no honest worker.
After performing at most $\nmalicious$ matches, the \master is guaranteed to identify all malicious workers.
The \master takes the encoded gradient of one worker identified as being honest from each group and sums up the group-wise results to obtain $\gestim=\tgrad$.
Algorithmic descriptions of the elimination tournament and a match between workers are given in \cref{alg:match} and \cref{alg:elimination_tournament} in \cref{app:algorithms}.

%% file: tikz/matchtree.tex
\begin{tikzpicture}
\def\angle{45}
\node {$\wpgrad{1}{\wind}+\wpgrad{2}{\wind}+\wpgrad{3}{\wind}+\wpgrad{4}{\wind}$} [sibling distance = 5cm,level distance=0.8cm]
    child {node (a) {$\wpgrad{1}{\wind}+\wpgrad{2}{\wind}$} [sibling distance = 2.5cm,level distance=1cm]
        child {node {$\wpgrad{1}{\wind}$}}
        child {node (y) {$\wpgrad{2}{\wind}$}}
    } 
    child {node (d) {$\wpgrad{3}{\wind}+\wpgrad{4}{\wind}$} [sibling distance = 2.5cm,level distance=0.8cm]
        child {node (z) {$\wpgrad{3}{\wind}$}}
        child {node {$\wpgrad{4}{\wind}$}}
    };
\end{tikzpicture}

%% file: Journal/sections/general_case.tex
In the following, we explain the extension of our scheme for the general case of $\nhonest\geq 1$. In this case, we also exploit the fact that there are $\nhonest$ honest workers that are guaranteed to agree in their responses. Therefore, the \master can eliminate all responses supported by less than $\nhonest$ workers. Furthermore, a response must be correct if it is supported by more than $\nmalicious$ workers.

\paragraph*{Extension of the example from \cref{sec:example}}
Consider the game presented in \cref{sec:example} for $4$ players, out of which $2$ players are liars. The goal of \playboss is again to identify the correct sum of integers, in this case $g_1+g_2+g_3+g_4$. First, we consider the naive question strategy by which \playbosspronoun simply asks for all claimed integer values. As explained in our first example, \playbosspronoun can easily find a $g_i$ on the value of which two players disagree and expose at least one liar by opening the corresponding envelope. That is, having $2$ liars, \playbosspronoun needs to open $2$ envelopes. However, \playbosspronoun realizes that the two honest players always agree on their responses in this setting. Considering the values in \cref{tab:example_generalization} when \playA is a liar and uses Strategy~A, then \playboss immediately exposes \playA and \playB as liars without having to open any envelope. Of course, \playA tries to avoid this and employs Strategy~B in the next game, aligning her values with \playB's. Clearly, \playboss needs to open an envelope to resolve the conflict. However, having opened a single envelope, \playbosspronoun can expose both liars at the same time. Thereby, \playbosspronoun can improve over the previous strategy. As suggested in our first example in \cref{sec:example}, \playboss can again reduce the communication by comparing partial sums between any two players who disagree on their total sum and, by this, identify an integer they disagree on.
\begin{table}[h]
\caption{Example for a game with $2$ liars out of $4$ players}
\label{tab:example_generalization}
\centering
\begin{minipage}{0.6\columnwidth}
\begin{tabularx}{\columnwidth}{l*{5}{X}}
        & $\exwpgrad{1}{\wind}$ & $\exwpgrad{2}{\wind}$ & $\exwpgrad{3}{\wind}$ & $\exwpgrad{4}{\wind}$ \\
        \toprule
        \playA (Strategy~A) & $1$ & $2$ & $4$ & $5$ \\
        \playA (Strategy~B) & $1$ & $3$ & $4$ & $5$ \\
        \playB & $1$ & $3$ & $4$ & $5$ \\
        \playC & $2$ & $3$ & $4$ & $5$ \\
        \playD & $2$ & $3$ & $4$ & $5$ \\
\end{tabularx}
\end{minipage}
\end{table}

\paragraph*{Formal construction}
From this example, we observe that, in general, the \master only needs to consider disjoint sets $\workerset_\ell \subseteq \range{\nworker}$ of workers that agree on their responses respectively, and which satisfy $\nhonest \leq \card{\workerset_\ell} \leq \nmalicious$.
Additionally, we leverage a little more communication to reduce the number of local gradient computations to at most $\compbound$. The intuition here is that the \master can pick one representative each from two contradicting sets $\workerset_{\ell_1}$ and $\workerset_{\ell_2}$ and run matches between the representatives. Note, however, that even if the \master identifies a representative of a group $\workerset_\ell$ as malicious by a local computation, this does not imply that every worker in $\workerset_\ell$ is malicious. For example, a malicious worker could return the same initial response as the honest workers, but when picked as a representative, they intentionally send wrong node labels. We overcome this issue by asking all workers in the sets $\workerset_{\ell_1}$ and $\workerset_{\ell_2}$ to either support or reject the challenging worker's claimed leaf label, which has been identified at the end of the match. The \master needs to run a local computation only if at least $\nhonest$ workers support the claim and at least $\nhonest$ workers reject the claim. If the set of supporting or rejecting workers is smaller than $\nhonest$, then the \master can mark all respective workers as malicious and, thus, resolve the match. Accordingly, if the \master needs to run a local computation, it can identify at least $\nhonest$ malicious workers. Namely, all those workers are exposed as liars who either supported a wrong claim from the challenging worker or rejected a correct one. This procedure is summarized by the elimination tournament in \cref{alg:elimination_tournament}. For the clarity of presentation, we only consider sets of size at least $\nhonest$ as inputs to \cref{alg:elimination_tournament}, since only these sets can provide an honest response, as explained above.

%% file: Journal/proofs/achievability_scheme.tex
\begin{IEEEproof}[Proof of \cref{thm:scheme}]
To show our scheme's correctness and figures of merit, we recall the following facts from before. The elimination tournament runs as long as there are contradicting responses among the non-eliminated workers. In each iteration of the elimination tournament, the main node eliminates at least one malicious worker (by either majority vote or local computation). The elimination terminates as soon as strictly less than $\nhonest$ malicious workers are left, in which case the resulting disagreements are resolved by majority vote.
W.l.o.g. we consider $1 \leq \nhonest \leq \nmalicious$ here. For $\nhonest > \nmalicious$, the elimination tournament will terminate immediately by majority vote about the full gradient. Furthermore, honest workers are never eliminated since they always respond with a correct value and never support an incorrect value. Having at least $\nhonest \geq 1$ honest workers in each fractional repetition group, the \master can recover the correct group result after the elimination tournament is terminated, and hence, for the output, it always holds that $\gestim=\tgrad$. This shows that our scheme is a valid \BGC scheme. In the remainder, we derive achievable values for the tuple (\nround, \localcomp, \replfact, \commoh).

We start with the number of locally computed gradients $\localcomp$.
As explained above, each local computation of a partial gradient eliminates at least $\nhonest$ malicious workers in our scheme. The procedure halts when there are no more discrepancies among the initial responses of the non-eliminated workers, which is at the latest when all $\nmalicious$ malicious workers are identified.
Therefore, the number of gradients computed locally is
\ifarxiv \else 

\vspace{-1em}
\fi
\begin{equation*}
  \localcomp \leq \compbound.
\end{equation*}

Next, we analyze the communication overhead $\commoh$.
The communication is determined by \begin{enumerate*}[label=\arabic*)] \item the number of matches; and \item the communication in each round. \end{enumerate*}
The number of matches depends on the number of local computations the \master needs to run. Suppose that the \master can identify all malicious workers with $\localcomp$ local computations in total, where $0 \leq \localcomp \leq \compbound$. As explained before, the protocol runs a local computation only if at least $\nhonest$ (malicious) workers support an erroneous partial gradient value or reject an honest partial gradient value. This implies that every local computation eliminates at least $\nhonest$ malicious workers, and the number of unidentified malicious workers after every local computation reduces accordingly. Hence, if there are any $\localcomp$ matches after which a local computation is run, the number of malicious workers that are not eliminated after the matches is at most $\nmalicious - \localcomp \nhonest$.
Note that by assumption on the chosen value for $\localcomp$, these malicious workers can be eliminated without any additional local computation.
By design of the scheme, every match identifies and eliminates at least one malicious worker.%
\footnote{The adversarial strategy that maximizes the communication overhead is to align all malicious worker responses within one fractional repetition group. The number of matches is now maximized if the malicious workers never commit to the representative's malicious value. This way, the protocol can only eliminate the representative of the malicious worker group.}
Eliminating $\localcomp\nhonest$ malicious workers by local computations, at most $\nmalicious-\localcomp\nhonest$ additional matches need to be run.
Therefore, the total number of matches is never higher than $\localcomp + (\nmalicious - \localcomp \nhonest)$ if $\localcomp \geq 1$. If $\localcomp = 0$, then the \master needs to run matches only until the number of unidentified malicious workers is reduced to $\nhonest-1$, hence requiring never more than $\nmalicious-(\nhonest-1)$ matches. In total, we can bound the number of matches by
\ifarxiv \else 

\vspace{-1em}
\fi
\begin{equation*}
  \nmalicious - \bar{\localcomp} (\nhonest-1).
\end{equation*}
with $\bar{\localcomp}=\max\{1,\localcomp\}$.
From the adversary's perspective, for $\nhonest>1$, the malicious workers can force more local computations only at the expense of reducing the number of matches.

The communication load during each match depends on the height of the match tree. In a match, a binary tree of height $\left\lceil \log_2\left(\frac{\ngrad}{\ngroup}\right) \right\rceil$ is traversed on a particular path. For each node on this path (excluding the root), the challenging worker sends a partial sum of gradients, which is one symbol from $\galpha$. The voting worker sends one bit for each node on this path. Hence, the accumulated communication load due to the matches is
\ifarxiv \else 

\vspace{-1em}
\fi
\begin{equation}
  \left( \nmalicious - \bar{\localcomp} (\nhonest-1) \right) \left\lceil \log_2\left(\frac{\ngrad}{\ngroup}\right) \right\rceil \left( \log_2\card{\galpha} + 1 \right).
  \label{eq:commoh_matches}
\end{equation}

After each match, the final voting round causes an additional communication load as follows. Every worker of the competing worker subsets (except for the two matching workers) transmits one bit, indicating whether or not the worker commits to the challenging worker's proposed value. The communication load caused by committing is maximized by maximizing the number of involved workers per voting round. This is achieved if the malicious workers group into a single large group. Furthermore, accumulated over all voting rounds, the number of involved workers depends on the point in the elimination tournament when the malicious workers force a local computation, after which their number is reduced by $\nhonest$. The number of involved workers accumulated over all voting rounds is maximized by forcing the local computations only for the last $\localcomp$ matches. In this case, the accumulated number of bits used for the voting rounds is
\ifarxiv
  \begin{align*}
    \underbrace{\sum_{k=1}^{\nmalicious - \localcomp\nhonest}\left( \nmalicious+\nhonest-2-(k-1) \right)}_{\text{matches concluded without local computations}} +& \\
    \underbrace{\sum_{k=1}^{\localcomp}\left( \nhonest+\localcomp\nhonest-2-\nhonest(k-1) \right)}_{\text{matches with local computations}}&
  \end{align*}
\else
  \vspace{-1em}

  \begin{align*}
    \underbrace{\sum_{k=1}^{\nmalicious - \localcomp\nhonest}\left( \nmalicious+\nhonest-2-(k-1) \right)}_{\text{matches concluded without local computations}} \!\!\!\!+ \,\,\underbrace{\sum_{k=1}^{\localcomp}\left( \nhonest+\localcomp\nhonest-2-\nhonest(k-1) \right)}_{\text{matches with local computations}} 
  \end{align*}
\fi
if $\localcomp \geq 1$ and $\sum_{k=1}^{\nmalicious-\nhonest+1} \nmalicious+\nhonest-2-(k-1)$ otherwise. These two expressions can be jointly written as
\ifarxiv \else 

\vspace{-1em}
\fi
\begin{equation}
  \left( \nmalicious-\bar{\localcomp}(\nhonest-1) \right) \frac{\nmalicious+(\bar{\localcomp}+2)\nhonest-3}{2} - \bar{\localcomp} \frac{\nmalicious-\nhonest + 1}{2}.
  \label{eq:commoh_voting}
\end{equation}
Adding \cref{eq:commoh_matches} and \cref{eq:commoh_voting} yields the upper bound on $\commoh$ stated in \cref{thm:scheme}.
We finally note that the communication overhead is never higher if the adversary chooses to corrupt workers from multiple fractional repetition groups. The reason is that fractional repetition groups are processed sequentially by the \master. As soon as the first erroneous fractional repetition groups with a particular number $\sigma$ of malicious workers have been resolved, the effective number of malicious workers reduces to $\nmalicious^\prime=\nmalicious-\sigma$. The effective minimum number of honest workers per fractional repetition group increases to $\nhonest^\prime=\nhonest+\sigma$. By our former derivation of the communication overhead per fractional repetition group, it can be shown that the communication overhead is maximized if all malicious workers are located in the same fractional repetition group.

In total, we obtain
  \ifarxiv
    \begin{equation}
      \label{eq:commoh_scheme}
      \begin{split}
        \commoh \leq&
        \left( \nmalicious-\bar{\localcomp}(\nhonest-1) \right)
        \cdot \Big(
          \left( 1 + \left\lceil \log_2{\card{\galpha}} \right\rceil \right) \left\lceil \log_2\left( \frac{\ngrad}{\ngroup}\right) \right\rceil \\
                    &+ \frac{\nmalicious+(\bar{\localcomp}+2)\nhonest-3}{2} \Big) - \bar{\localcomp} \frac{\nmalicious-\nhonest + 1}{2}.
      \end{split}
    \end{equation}
  \else
    \vspace{-1em}
  
    \begin{equation}
      \label{eq:commoh_scheme}
      \begin{split}
        \commoh \leq&
        \left( \nmalicious-\bar{\localcomp}(\nhonest-1) \right)
        \cdot \left(
          \left( 1 + \left\lceil \log_2{\card{\galpha}} \right\rceil \right) \left\lceil \log_2\left( \frac{\ngrad}{\ngroup}\right) \right\rceil
          + \frac{\nmalicious+(\bar{\localcomp}+2)\nhonest-3}{2}
        \right) - \bar{\localcomp} \frac{\nmalicious-\nhonest + 1}{2}.
      \end{split}
    \end{equation}
  \fi
 
Finally, we analyze the number of communication rounds $\nround$. The number of communication rounds in a match is again upper bounded by the height of the tree, i.e., $\left\lceil \log_2\left(\frac{\ngrad}{\ngroup}\right) \right\rceil$, multiplied by a factor of $2$, since each descent step in the match tree requires one round for the challenging worker to propose a partial sum value and one round for the voting worker to react to it. In the worst case, there is one additional voting round after every match. In total, the resolution of a match requires at most $2\left\lceil \log_2\left(\frac{\ngrad}{\ngroup}\right) \right\rceil + 1$ communication rounds.
As explained before, there can be up to $\nmalicious - \bar{\localcomp} (\nhonest-1)$ matches. Again, if all malicious workers are in the same group, and if there are only two consistent worker subsets in this group of size $\nhonest$ and $\nmalicious$, respectively, all of these matches must be executed sequentially.
Resolving all conflicts, thus, requires
\ifarxiv \else 

\vspace{-1em}
\fi
\begin{equation*}
  \nround
  \leq
  \left( \nmalicious - \bar{\localcomp} (\nhonest-1) \right)
  \left( 2\left\lceil \log_2\left(\frac{\ngrad}{\ngroup}\right) \right\rceil + 1 \right).
\end{equation*}
\end{IEEEproof}

\begin{IEEEproof}[Large datasets (\cref{cor:achievable_commoh})]
  For large dataset sizes $\ngrad$, the communication overhead of our scheme is asymptotically bounded as
  \ifarxiv \else 
  
  \vspace{-1em}
  \fi
  \begin{equation*}
    \commoh \leq
    \left( \nmalicious-\bar{\localcomp}(\nhonest-1) \right)
    \left( 1+ \left\lceil \log_2{\card{\galpha}} \right\rceil \right)
    \left\lceil \log_2\left( \frac{\ngrad}{\ngroup}\right) \right\rceil.
  \end{equation*}
\end{IEEEproof}
\begin{IEEEproof}[Extension to stragglers (\cref{cor:stagglers})]
  For the asymptotic communication overhead, i.e., for large $\ngrad$, the communication is dominated by the communication of the matches (compare \cref{eq:commoh_scheme}). That is, the communication overhead is determined by the transmission of at most $\left\lceil \log_2\left( \frac{\ngrad}{\ngroup}\right) \right\rceil$ gradients and the same number of voting bits per match. In the case of $\nstraggle$ stragglers, the effective number of honest workers in a fractional repetition group can reduce to $\nhonest-\nstraggle$. Therefore, $\left( \nmalicious-\bar{\localcomp}(\nhonest-\nstraggle-1) \right)$ matches need to be performed in the worst case to resolve the conflicts. Hence, we obtain
  \ifarxiv \else 

  \vspace{-1em}
  \fi
  \begin{equation*}
    \commoh \!\leq\!
    \left( \nmalicious-\bar{\localcomp}(\nhonest-\nstraggle-1) \right)
    \left( 1\!+\!\left\lceil \log_2{\card{\galpha}} \right\rceil \right)
    \left\lceil \log_2\left( \frac{\ngrad}{\ngroup}\right) \right\rceil,
  \end{equation*}
  since the communication load for the matches dominates the communication in voting rounds.
\end{IEEEproof}

%% file: tikz/comm_vs_localcomp_updated2.tex
\begin{tikzpicture}
\begin{axis}[
legend cell align={left}, 
legend columns={1}, 
legend style={color={rgb,1:red,0.0;green,0.0;blue,0.0}, draw opacity={1.0}, line width={1}, solid, fill={rgb,1:red,1.0;green,1.0;blue,1.0}, fill opacity={1.0}, text opacity={1.0}, font={{\fontsize{8 pt}{10.4 pt}\selectfont}}, text={rgb,1:red,0.0;green,0.0;blue,0.0}, cells={anchor={center}}, at={(0.98, 0.98)}, anchor={north east}}, 
axis background/.style={fill={rgb,1:red,1.0;green,1.0;blue,1.0}, opacity={1.0}}, 
anchor={north west}, 
width={93mm}, 
height={50mm}, 
xlabel={local computations $c$}, 
x tick style={color={rgb,1:red,0.0;green,0.0;blue,0.0}, opacity={1.0}}, 
x tick label style={color={rgb,1:red,0.0;green,0.0;blue,0.0}, opacity={1.0}, rotate={0}}, 
xmin={-0.8}, 
xmax={10.5}, 
xtick={{10,5,3,2,2,1,1,1,1,1,0}}, 
xticklabels={{$10$,$5$,$3$,$2$,$2$,$1$,$1$,$1$,$1$,$1$,$0$\\(DRACO)}}, 
xtick align={inside}, 
xticklabel style={align=center},
axis x line*={left}, 
ylabel={total worker to main node\\communication [\si{\giga\byte}]}, 
ylabel style={align=center},
ymajorgrids={true}, 
ymin={0},
ymax={402.890145778656}, 
axis y line*={left}, 
y axis line style={color={rgb,1:red,0.0;green,0.0;blue,0.0}, draw opacity={1.0}, line width={1}, solid}, 
colorbar={false}]
 \addplot[color={rgb,1:red,0.0;green,0.0;blue,0.0}, name path={da9859dc-9801-4f28-8103-cdd0ab16e5ff}, area legend, fill={rgb,1:red,0.0667;green,0.4392;blue,0.6667}, fill opacity={1.0}, draw opacity={1.0}, line width={1}, solid]
        table[row sep={\\}]
        {
            \\
            -0.4  391.155481338501  \\
            -0.4  0.0  \\
            0.4  0.0  \\
            0.4  391.155481338501  \\
            -0.4  391.155481338501  \\
        }
        ;
    \addplot[color={rgb,1:red,0.0;green,0.0;blue,0.0}, name path={2c985cfe-24f1-4bdd-8479-e66bb8f910a2}, area legend, fill={rgb,1:red,0.0667;green,0.4392;blue,0.6667}, fill opacity={1.0}, draw opacity={1.0}, line width={1}, solid, forget plot]
        table[row sep={\\}]
        {
            \\
            0.6  298.02322620525956  \\
            0.6  0.0  \\
            1.4  0.0  \\
            1.4  298.02322620525956  \\
            0.6  298.02322620525956  \\
        }
        ;
    \addplot[color={rgb,1:red,0.0;green,0.0;blue,0.0}, name path={2c985cfe-24f1-4bdd-8479-e66bb8f910a2}, area legend, fill={rgb,1:red,0.0667;green,0.4392;blue,0.6667}, fill opacity={1.0}, draw opacity={1.0}, line width={1}, solid, forget plot]
        table[row sep={\\}]
        {
            \\
            1.6  260.7703227382153  \\
            1.6  0.0  \\
            2.4  0.0  \\
            2.4  260.7703227382153  \\
            1.6  260.7703227382153  \\
        }
        ;
    \addplot[color={rgb,1:red,0.0;green,0.0;blue,0.0}, name path={2c985cfe-24f1-4bdd-8479-e66bb8f910a2}, area legend, fill={rgb,1:red,0.0667;green,0.4392;blue,0.6667}, fill opacity={1.0}, draw opacity={1.0}, line width={1}, solid, forget plot]
        table[row sep={\\}]
        {
            \\
            2.6  242.14387123286724  \\
            2.6  0.0  \\
            3.4  0.0  \\
            3.4  242.14387123286724  \\
            2.6  242.14387123286724  \\
        }
        ;
    \addplot[color={rgb,1:red,0.0;green,0.0;blue,0.0}, name path={2c985cfe-24f1-4bdd-8479-e66bb8f910a2}, area legend, fill={rgb,1:red,0.0667;green,0.4392;blue,0.6667}, fill opacity={1.0}, draw opacity={1.0}, line width={1}, solid, forget plot]
        table[row sep={\\}]
        {
            \\
            4.6  223.51742018014193  \\
            4.6  0.0  \\
            5.4  0.0  \\
            5.4  223.51742018014193  \\
            4.6  223.51742018014193  \\
        }
        ;
    \addplot[color={rgb,1:red,0.0;green,0.0;blue,0.0}, name path={2c985cfe-24f1-4bdd-8479-e66bb8f910a2}, area legend, fill={rgb,1:red,0.0667;green,0.4392;blue,0.6667}, fill opacity={1.0}, draw opacity={1.0}, line width={1}, solid, forget plot]
        table[row sep={\\}]
        {
            \\
            9.6  204.89097093231976  \\
            9.6  0.0  \\
            10.4  0.0  \\
            10.4  204.89097093231976  \\
            9.6  204.89097093231976  \\
        }
        ;
    \addplot[color={rgb,1:red,0.0667;green,0.4392;blue,0.6667}, name path={bc1281c6-5722-4353-8df1-840e886f6d2e}, only marks, draw opacity={1.0}, line width={0}, solid, mark={*}, mark size={0.0 pt}, mark repeat={1}, mark options={color={rgb,1:red,0.0;green,0.0;blue,0.0}, draw opacity={0.0}, fill={rgb,1:red,0.0667;green,0.4392;blue,0.6667}, fill opacity={0.0}, line width={0.75}, rotate={0}, solid}, forget plot]
        table[row sep={\\}]
        {
            \\
            0.0  391.155481338501  \\
            1.0  298.02322620525956  \\
            2.0  260.7703227382153  \\
            3.0  242.14387123286724  \\
            5.0  223.51742018014193  \\
            10.0  204.89097093231976  \\
        }
        ;
\end{axis}
\end{tikzpicture}

%% file: Journal/sections/achievability_vs_converse.tex
For fixed numbers of honest and malicious workers the worst case communication overhead of the proposed scheme decreases with the number of local computations forced by the malicious workers. The highest communication overhead can occur when no local computations need to be performed, i.e., $\localcomp=0$. The lowest communication overhead occurs when the maximum number of local computations needs to be performed, i.e., $\localcomp=\compbound$.
In the latter case, our scheme asymptotically reaches the lower bound on the communication overhead up to a constant factor (see \cref{cor:achievable_commoh}).
Typically, in distributed gradient descent applications the number of parameters $\graddim$ and the number of samples per group $\ngrad$ are very large, whereas the number of workers $\nworker$ and as a consequence $\nmalicious$, $\nhonest$ and $\ngroup$ are small by comparison.
For large numbers of samples $\ngrad$, the ratio $\commoh[achieve]/\commoh[bound]$ tends to
\ifarxiv \else 

\vspace{-1em}
\fi
\begin{align}
  \lim_{\ngrad \to \infty} \frac{\commoh[achieve]}{\commoh[bound]} &= \left( 1+ \left\lceil \log_2\card{\galpha} \right\rceil \right) \frac{\nmalicious-\bar{\localcomp}(\nhonest-1)}{\lfloor \nmalicious / \nhonest 
  \rfloor}
  \nonumber
  \\
                                                                   &= \left( 1+ \left\lceil \log_2\card{\galpha} \right\rceil \right) \left( 1+ \frac{\nmalicious \operatorname{mod} \nhonest}{\nmalicious \operatorname{div} \nhonest} \right).
                                                                   \label{eq:conv_limit}
  \end{align}
The convergence behavior can be observed in \cref{fig:convergence}.
\begin{figure}[ht]
    \centering
    \ifarxiv
    \resizebox{0.98\linewidth}{!}{ \input{tikz/kappa_relative_vs_p_updated2}}
    \else
    \resizebox{0.5\linewidth}{!}{ \input{tikz/kappa_relative_vs_p_updated2}}
    \fi
    \caption{Convergence of the ratio $\frac{\commoh[achieve]}{\commoh[bound]}$ to the limit given in $\eqref{eq:conv_limit}$ for large numbers of samples.
      The parameters are $\nworker=10$, $\ngroup=10$, $\card{\galpha}=2^{16}$.
    For $\nmalicious \in \{5,9\}$ the limits as in \eqref{eq:conv_limit} give the same value.
    }
    \label{fig:convergence}
\end{figure}

Note that depending on the particular alphabet the communication overhead of our scheme can be slightly improved as stated in the following.
\begin{remark}[Compression beyond the alphabet size]
If for every pair of elements $a, b \in \galpha$ there exists a
function $f: \galpha \to \mathcal{B}$, that maps from $\galpha$ to a
smaller alphabet $\mathcal{B}$, such that $f(a) \neq f(b)$ and an operation $\oplus$ with the property $\forall c, d, e, g \in \galpha: f(c+d) \neq f(e+g) \implies f(c) \neq f(d)\text{ or }f(e) \neq f(g)$, then 
our scheme can be improved. Namely, each node label can be communication with $\log_2\card{\mathcal{B}}$ bits instead of $\log_2\card{\galpha}$ bits.
At the start of each match, the main node would choose not only the index $\compind$ but also the appropriate compression function $f$ and communicate it to the two workers. They then use $f$ to compress their transmitted symbols during the match. 
\end{remark}

%% file: tikz/kappa_relative_vs_p_updated2.tex
\begin{tikzpicture}
\begin{axis}[
  legend cell align={left}, 
  legend columns={3}, 
  title style={at={{(0.1,1)}}, anchor={south}, color={rgb,1:red,0.0;green,0.0;blue,0.0}, draw opacity={1.0}, rotate={0.0}}, 
    legend style={
            /tikz/column 2/.style={
                column sep=3pt,
            },
      font={{\fontsize{8 pt}{10.4 pt}\selectfont}},
      cells={anchor={center}},
      at={(0,1.05)},
      anchor=south west
      }, 
  axis background/.style={fill={rgb,1:red,1.0;green,1.0;blue,1.0}, opacity={1.0}}, 
  anchor={north west}, 
  xshift={1.0mm}, 
  yshift={-1.0mm}, 
  width={90mm}, 
  height={50mm}, 
  xlabel={number of samples $\ngrad$}, 
  x tick style={color={rgb,1:red,0.0;green,0.0;blue,0.0}, opacity={1.0}}, 
  x tick label style={color={rgb,1:red,0.0;green,0.0;blue,0.0}, opacity={1.0}, rotate={0}}, 
  xmode={log}, 
  log basis x={10}, 
  xmajorgrids={true}, 
  xmin={100}, 
  xmax={1e10}, 
  x tick label style={ color={rgb,1:red,0.0;green,0.0;blue,0.0}, draw opacity={1.0}, rotate={0.0}}, 
  x grid style={color={rgb,1:red,0.0;green,0.0;blue,0.0}, draw opacity={0.1}, line width={0.5}, solid}, 
  axis x line*={left}, 
  x axis line style={color={rgb,1:red,0.0;green,0.0;blue,0.0}, draw opacity={1.0}, line width={1}, solid}, 
  scaled y ticks={false}, 
  ylabel={$\commoh[achieve] / \commoh[bound]$}, 
  y tick style={color={rgb,1:red,0.0;green,0.0;blue,0.0}, opacity={1.0}}, 
  y tick label style={color={rgb,1:red,0.0;green,0.0;blue,0.0}, opacity={1.0}, rotate={0}}, 
  ymode={log}, 
  log basis y={10}, 
  ymajorgrids={true}, 
  ytick align={inside}, 
  y grid style={color={rgb,1:red,0.0;green,0.0;blue,0.0}, draw opacity={0.1}, line width={0.5}, solid}, 
  axis y line*={left}, 
  y axis line style={color={rgb,1:red,0.0;green,0.0;blue,0.0}, draw opacity={1.0}, line width={1}, solid}, 
  colorbar={false}]
    
  \addplot[color={rgb,1:red,0.0667;green,0.4392;blue,0.6667}, 
  name path={1eeefe02-17e6-4c94-a00a-326681316fae}, 
  draw opacity={1.0}, 
  line width={1}, solid]
        table[row sep={\\}]
        {
            \\
            100.0  27.13845230552393  \\
            126.0  25.2137623758321  \\
            158.0  26.84842886632568  \\
            200.0  25.168725961244768  \\
            251.0  23.74425735269964  \\
            316.0  25.187595758000082  \\
            398.0  23.8968813393522  \\
            501.0  22.737289638426624  \\
            631.0  24.03254002955749  \\
            794.0  22.974618240179105  \\
            1000.0  22.003393296075714  \\
            1259.0  23.17583515810262  \\
            1585.0  22.275367342204053  \\
            1995.0  21.44344615176975  \\
            2512.0  22.510347510399026  \\
            3162.0  21.72852714388933  \\
            3981.0  20.99877725685209  \\
            5012.0  21.977178818388065  \\
            6310.0  21.285888181878693  \\
            7943.0  20.637249626593018  \\
            10000.0  21.539813157474846  \\
            12589.0  20.92105972667233  \\
            15849.0  20.336793897355466  \\
            19953.0  21.174137989946782  \\
            25119.0  20.614231255783128  \\
            31623.0  20.08315788273786  \\
            39811.0  20.863872205438472  \\
            50119.0  20.35273806725782  \\
            63096.0  19.866057323808604  \\
            79433.0  20.59717418353592  \\
            100000.0  20.127141885967728  \\
            125893.0  19.678076327660982  \\
            158489.0  20.365446928872917  \\
            199526.0  19.930484507973283  \\
            251189.0  19.51371212900521  \\
            316228.0  20.16220842045208  \\
            398107.0  19.75752462911729  \\
            501187.0  19.368765347254477  \\
            630957.0  19.982518635739577  \\
            794328.0  19.604220225713483  \\
            1.0e6  19.2399795240163  \\
            1.258925e6  19.822497570973272  \\
            1.584893e6  19.467397373824493  \\
            1.995262e6  19.12479635557424  \\
            2.511886e6  19.679083736630858  \\
            3.162278e6  19.34453246593242  \\
            3.981072e6  19.02116655013364  \\
            5.011872e6  19.54981856109094  \\
            6.309573e6  19.233593274895703  \\
            7.943282e6  18.927435265195584  \\
            1.0e7  19.432706291640084  \\
            1.2589254e7  19.132922771857654  \\
            1.5848932e7  18.842248095304456  \\
            1.9952623e7  19.3261098988419  \\
            2.5118864e7  19.041158484852158  \\
            3.1622777e7  18.764487811518393  \\
            3.9810717e7  19.228673674216775  \\
            5.0118723e7  18.957168447121582  \\
            6.3095734e7  18.69322367058508  \\
            7.9432823e7  19.13926549843139  \\
            1.0e8  18.88000483057183  \\
            1.25892541e8  18.62767419368251  \\
            1.58489319e8  19.056932406177296  \\
            1.99526231e8  18.80886788895002  \\
            2.51188643e8  18.567178507378685  \\
            3.16227766e8  18.98086655914894  \\
            3.98107171e8  18.74307808693781  \\
            5.01187234e8  18.51117383267841  \\
            6.30957344e8  18.910378589467683  \\
            7.94328235e8  18.682054376861608  \\
            1.0e9  18.459177969404543  \\
            1.258925412e9  18.84487665554034  \\
            1.584893192e9  18.625296943294217  \\
            1.995262315e9  18.410775360543063  \\
            2.511886432e9  18.78384979203553  \\
            3.16227766e9  18.572373500982295  \\
            3.981071706e9  18.36560597023359  \\
            5.011872336e9  18.726854543245636  \\
            6.309573445e9  18.522908260826306  \\
            7.943282347e9  18.32335630783538  \\
            1.0e10  18.673504172318392  \\
        }
        ;
    \addlegendentry {scheme $\nmalicious=9$}
    \addplot[color={rgb,1:red,0.6392;green,0.6745;blue,0.7255},  draw opacity={1.0}, line width={1}, solid]
        table[row sep={\\}]
        {
            \\
            100.0  30.635825013814177  \\
            126.0  29.05030786781474  \\
            158.0  31.40074957281149  \\
            200.0  29.900954013500076  \\
            251.0  28.586103526241743  \\
            316.0  30.635156324941114  \\
            398.0  29.379912091146043  \\
            501.0  28.226483267887595  \\
            631.0  30.055615592821656  \\
            794.0  28.963809182074588  \\
            1000.0  27.944875997368097  \\
            1259.0  29.599419661116215  \\
            1585.0  28.628369771185724  \\
            1995.0  27.71986513085635  \\
            2512.0  29.228298263300722  \\
            3162.0  28.355998769345792  \\
            3981.0  27.53358283460676  \\
            5012.0  28.92029735115909  \\
            6310.0  28.127631896754743  \\
            7943.0  27.37773310447834  \\
            10000.0  28.660487242944345  \\
            12589.0  27.93483710053311  \\
            15849.0  27.244900480747788  \\
            19953.0  28.438187985164816  \\
            25119.0  27.768996279916877  \\
            31623.0  27.130536444000793  \\
            39811.0  28.24598577407882  \\
            50119.0  27.625121656448197  \\
            63096.0  27.030965096615727  \\
            79433.0  28.078062711337157  \\
            100000.0  27.499074176599134  \\
            125893.0  26.943468229189012  \\
            158489.0  27.930098932617764  \\
            199526.0  27.387708968167303  \\
            251189.0  26.865980102759817  \\
            316228.0  27.798718559572276  \\
            398107.0  27.288603925075773  \\
            501187.0  26.79687121147223  \\
            630957.0  27.68130170760639  \\
            794328.0  27.199843745105213  \\
            1.0e6  26.734847340386047  \\
            1.258925e6  27.575723991381146  \\
            1.584893e6  27.119887515061077  \\
            1.995262e6  26.678876737085396  \\
            2.511886e6  27.48028307813133  \\
            3.162278e6  27.04748701077687  \\
            3.981072e6  26.62811264242354  \\
            5.011872e6  27.39358580734786  \\
            6.309573e6  26.98162024756761  \\
            7.943282e6  26.581861962718943  \\
            1.0e7  27.314483182176872  \\
            1.2589254e7  26.921439908236405  \\
            1.5848932e7  26.53954757779052  \\
            1.9952623e7  27.242019183307995  \\
            2.5118864e7  26.866240500527468  \\
            3.1622777e7  26.500687758958605  \\
            3.9810717e7  27.175391612943386  \\
            5.0118723e7  26.815428102229237  \\
            6.3095734e7  26.46487602751596  \\
            7.9432823e7  27.113922674710917  \\
            1.0e8  26.768499713490957  \\
            1.25892541e8  26.43176721235468  \\
            1.58489319e8  27.05703555687708  \\
            1.99526231e8  26.725026411232086  \\
            2.51188643e8  26.40106647086048  \\
            3.16227766e8  27.004236394573343  \\
            3.98107171e8  26.684640152154444  \\
            5.01187234e8  26.372520297778703  \\
            6.30957344e8  26.955099824035827  \\
            7.94328235e8  26.647023406206195  \\
            1.0e9  26.34590957443921  \\
            1.258925412e9  26.9092575132062  \\
            1.584893192e9  26.611900730317664  \\
            1.995262315e9  26.32104391414456  \\
            2.511886432e9  26.866388924617358  \\
            3.16227766e9  26.579032039244154  \\
            3.981071706e9  26.297757112022826  \\
            5.011872336e9  26.826213796208357  \\
            6.309573445e9  26.548207106704965  \\
            7.943282347e9  26.27590341993949  \\
            1.0e10  26.788486011398923  \\
        }
        ;
    \addlegendentry {scheme $\nmalicious=7$}
    \addplot[color={rgb,1:red,0.9882;green,0.4902;blue,0.0431}, name path={44590de5-03d4-4a3a-937b-5ede02b3c2bf}, draw opacity={1.0}, line width={1}, solid]
        table[row sep={\\}]
        {
            \\
            100.0  19.115404724662806  \\
            126.0  18.201935624261832  \\
            158.0  19.715816364824466  \\
            200.0  18.83865897328504  \\
            251.0  18.064255564235392  \\
            316.0  19.38875855258386  \\
            398.0  18.641541930285438  \\
            501.0  17.951386054631378  \\
            631.0  19.136706210342535  \\
            794.0  18.47815724571241  \\
            1000.0  17.861113076062885  \\
            1259.0  18.935600625622513  \\
            1585.0  18.343847591648263  \\
            1995.0  17.788443901278214  \\
            2512.0  18.769997214895785  \\
            3162.0  18.234015549736018  \\
            3981.0  17.72736048654183  \\
            5012.0  18.63126024379717  \\
            6310.0  18.140878001206747  \\
            7943.0  17.675933814976137  \\
            10000.0  18.513344733334844  \\
            12589.0  18.06183835893616  \\
            15849.0  17.63174905784721  \\
            19953.0  18.41179949670951  \\
            25119.0  17.993374238288048  \\
            31623.0  17.593518862622606  \\
            39811.0  18.32355265824539  \\
            50119.0  17.933692722617476  \\
            63096.0  17.56007639360241  \\
            79433.0  18.24610022488071  \\
            100000.0  17.881181742440482  \\
            125893.0  17.53056489505894  \\
            158489.0  18.177583565442006  \\
            199526.0  17.834608985857862  \\
            251189.0  17.504334784949812  \\
            316228.0  18.11653140784698  \\
            398107.0  17.793023549158374  \\
            501187.0  17.480865281350674  \\
            630957.0  18.061800478221947  \\
            794328.0  17.755667927072004  \\
            1.0e6  17.45973974851091  \\
            1.258925e6  18.012450979698244  \\
            1.584893e6  17.72192731483549  \\
            1.995262e6  17.440626922703288  \\
            2.511886e6  17.967728075647553  \\
            3.162278e6  17.6913011566105  \\
            3.981072e6  17.42325117938655  \\
            5.011872e6  17.927010267508244  \\
            6.309573e6  17.663377766281418  \\
            7.943282e6  17.407386753683053  \\
            1.0e7  17.889782599459455  \\
            1.2589254e7  17.637813840560348  \\
            1.5848932e7  17.392844188928578  \\
            1.9952623e7  17.855614819271295  \\
            2.5118864e7  17.614322732224924  \\
            3.1622777e7  17.37946507032436  \\
            3.9810717e7  17.824144481520253  \\
            5.0118723e7  17.59266209143998  \\
            6.3095734e7  17.367115141326618  \\
            7.9432823e7  17.795064306461192  \\
            1.0e8  17.572625996884902  \\
            1.25892541e8  17.355679998249546  \\
            1.58489319e8  17.7681119406524  \\
            1.99526231e8  17.554038303656903  \\
            2.51188643e8  17.345061655463567  \\
            3.16227766e8  17.743062097419397  \\
            3.98107171e8  17.536747420827048  \\
            5.01187234e8  17.335175611730534  \\
            6.30957344e8  17.71972019997716  \\
            7.94328235e8  17.52062221924731  \\
            1.0e9  17.325948639326917  \\
            1.258925412e9  17.6979173273597  \\
            1.584893192e9  17.505548661134984  \\
            1.995262315e9  17.31731695485082  \\
            2.511886432e9  17.67750612819348  \\
            3.16227766e9  17.491427116518118  \\
            3.981071706e9  17.30922475058745  \\
            5.011872336e9  17.658357477651816  \\
            6.309573445e9  17.47817015638439  \\
            7.943282347e9  17.30162298313485  \\
            1.0e10  17.6403577459093  \\
        }
        ;
    \addlegendentry {scheme $\nmalicious=5$}
    \addplot[color={rgb,1:red,0.0667;green,0.4392;blue,0.6667}, name path={233b7f50-a717-44f1-9f9e-0bb573127f34}, draw opacity={1.0}, line width={1}, thick, dashed]
        table[row sep={\\}]
        {
            \\
            100.0  17.0  \\
            126.0  17.0  \\
            158.0  17.0  \\
            200.0  17.0  \\
            251.0  17.0  \\
            316.0  17.0  \\
            398.0  17.0  \\
            501.0  17.0  \\
            631.0  17.0  \\
            794.0  17.0  \\
            1000.0  17.0  \\
            1259.0  17.0  \\
            1585.0  17.0  \\
            1995.0  17.0  \\
            2512.0  17.0  \\
            3162.0  17.0  \\
            3981.0  17.0  \\
            5012.0  17.0  \\
            6310.0  17.0  \\
            7943.0  17.0  \\
            10000.0  17.0  \\
            12589.0  17.0  \\
            15849.0  17.0  \\
            19953.0  17.0  \\
            25119.0  17.0  \\
            31623.0  17.0  \\
            39811.0  17.0  \\
            50119.0  17.0  \\
            63096.0  17.0  \\
            79433.0  17.0  \\
            100000.0  17.0  \\
            125893.0  17.0  \\
            158489.0  17.0  \\
            199526.0  17.0  \\
            251189.0  17.0  \\
            316228.0  17.0  \\
            398107.0  17.0  \\
            501187.0  17.0  \\
            630957.0  17.0  \\
            794328.0  17.0  \\
            1.0e6  17.0  \\
            1.258925e6  17.0  \\
            1.584893e6  17.0  \\
            1.995262e6  17.0  \\
            2.511886e6  17.0  \\
            3.162278e6  17.0  \\
            3.981072e6  17.0  \\
            5.011872e6  17.0  \\
            6.309573e6  17.0  \\
            7.943282e6  17.0  \\
            1.0e7  17.0  \\
            1.2589254e7  17.0  \\
            1.5848932e7  17.0  \\
            1.9952623e7  17.0  \\
            2.5118864e7  17.0  \\
            3.1622777e7  17.0  \\
            3.9810717e7  17.0  \\
            5.0118723e7  17.0  \\
            6.3095734e7  17.0  \\
            7.9432823e7  17.0  \\
            1.0e8  17.0  \\
            1.25892541e8  17.0  \\
            1.58489319e8  17.0  \\
            1.99526231e8  17.0  \\
            2.51188643e8  17.0  \\
            3.16227766e8  17.0  \\
            3.98107171e8  17.0  \\
            5.01187234e8  17.0  \\
            6.30957344e8  17.0  \\
            7.94328235e8  17.0  \\
            1.0e9  17.0  \\
            1.258925412e9  17.0  \\
            1.584893192e9  17.0  \\
            1.995262315e9  17.0  \\
            2.511886432e9  17.0  \\
            3.16227766e9  17.0  \\
            3.981071706e9  17.0  \\
            5.011872336e9  17.0  \\
            6.309573445e9  17.0  \\
            7.943282347e9  17.0  \\
            1.0e10  17.0  \\
        }
        ;
      \addlegendentry {limit $\nmalicious=9$}
    \addplot[color={rgb,1:red,0.6392;green,0.6745;blue,0.7255}, name path={f75f2af8-a259-464f-ac67-414608860373}, draw opacity={1.0}, line width={1}, thick, dashed]
        table[row sep={\\}]
        {
            \\
            100.0  25.5  \\
            126.0  25.5  \\
            158.0  25.5  \\
            200.0  25.5  \\
            251.0  25.5  \\
            316.0  25.5  \\
            398.0  25.5  \\
            501.0  25.5  \\
            631.0  25.5  \\
            794.0  25.5  \\
            1000.0  25.5  \\
            1259.0  25.5  \\
            1585.0  25.5  \\
            1995.0  25.5  \\
            2512.0  25.5  \\
            3162.0  25.5  \\
            3981.0  25.5  \\
            5012.0  25.5  \\
            6310.0  25.5  \\
            7943.0  25.5  \\
            10000.0  25.5  \\
            12589.0  25.5  \\
            15849.0  25.5  \\
            19953.0  25.5  \\
            25119.0  25.5  \\
            31623.0  25.5  \\
            39811.0  25.5  \\
            50119.0  25.5  \\
            63096.0  25.5  \\
            79433.0  25.5  \\
            100000.0  25.5  \\
            125893.0  25.5  \\
            158489.0  25.5  \\
            199526.0  25.5  \\
            251189.0  25.5  \\
            316228.0  25.5  \\
            398107.0  25.5  \\
            501187.0  25.5  \\
            630957.0  25.5  \\
            794328.0  25.5  \\
            1.0e6  25.5  \\
            1.258925e6  25.5  \\
            1.584893e6  25.5  \\
            1.995262e6  25.5  \\
            2.511886e6  25.5  \\
            3.162278e6  25.5  \\
            3.981072e6  25.5  \\
            5.011872e6  25.5  \\
            6.309573e6  25.5  \\
            7.943282e6  25.5  \\
            1.0e7  25.5  \\
            1.2589254e7  25.5  \\
            1.5848932e7  25.5  \\
            1.9952623e7  25.5  \\
            2.5118864e7  25.5  \\
            3.1622777e7  25.5  \\
            3.9810717e7  25.5  \\
            5.0118723e7  25.5  \\
            6.3095734e7  25.5  \\
            7.9432823e7  25.5  \\
            1.0e8  25.5  \\
            1.25892541e8  25.5  \\
            1.58489319e8  25.5  \\
            1.99526231e8  25.5  \\
            2.51188643e8  25.5  \\
            3.16227766e8  25.5  \\
            3.98107171e8  25.5  \\
            5.01187234e8  25.5  \\
            6.30957344e8  25.5  \\
            7.94328235e8  25.5  \\
            1.0e9  25.5  \\
            1.258925412e9  25.5  \\
            1.584893192e9  25.5  \\
            1.995262315e9  25.5  \\
            2.511886432e9  25.5  \\
            3.16227766e9  25.5  \\
            3.981071706e9  25.5  \\
            5.011872336e9  25.5  \\
            6.309573445e9  25.5  \\
            7.943282347e9  25.5  \\
            1.0e10  25.5  \\
        }
        ;
        \addlegendentry {limit $\nmalicious=7$}
    \addplot[color={rgb,1:red,0.9882;green,0.4902;blue,0.0431}, name path={6095e951-ac26-4dd8-b5f0-843e048439e3}, draw opacity={1.0}, line width={1}, thick, dashdotted]
        table[row sep={\\}]
        {
            \\
            100.0  17.0  \\
            126.0  17.0  \\
            158.0  17.0  \\
            200.0  17.0  \\
            251.0  17.0  \\
            316.0  17.0  \\
            398.0  17.0  \\
            501.0  17.0  \\
            631.0  17.0  \\
            794.0  17.0  \\
            1000.0  17.0  \\
            1259.0  17.0  \\
            1585.0  17.0  \\
            1995.0  17.0  \\
            2512.0  17.0  \\
            3162.0  17.0  \\
            3981.0  17.0  \\
            5012.0  17.0  \\
            6310.0  17.0  \\
            7943.0  17.0  \\
            10000.0  17.0  \\
            12589.0  17.0  \\
            15849.0  17.0  \\
            19953.0  17.0  \\
            25119.0  17.0  \\
            31623.0  17.0  \\
            39811.0  17.0  \\
            50119.0  17.0  \\
            63096.0  17.0  \\
            79433.0  17.0  \\
            100000.0  17.0  \\
            125893.0  17.0  \\
            158489.0  17.0  \\
            199526.0  17.0  \\
            251189.0  17.0  \\
            316228.0  17.0  \\
            398107.0  17.0  \\
            501187.0  17.0  \\
            630957.0  17.0  \\
            794328.0  17.0  \\
            1.0e6  17.0  \\
            1.258925e6  17.0  \\
            1.584893e6  17.0  \\
            1.995262e6  17.0  \\
            2.511886e6  17.0  \\
            3.162278e6  17.0  \\
            3.981072e6  17.0  \\
            5.011872e6  17.0  \\
            6.309573e6  17.0  \\
            7.943282e6  17.0  \\
            1.0e7  17.0  \\
            1.2589254e7  17.0  \\
            1.5848932e7  17.0  \\
            1.9952623e7  17.0  \\
            2.5118864e7  17.0  \\
            3.1622777e7  17.0  \\
            3.9810717e7  17.0  \\
            5.0118723e7  17.0  \\
            6.3095734e7  17.0  \\
            7.9432823e7  17.0  \\
            1.0e8  17.0  \\
            1.25892541e8  17.0  \\
            1.58489319e8  17.0  \\
            1.99526231e8  17.0  \\
            2.51188643e8  17.0  \\
            3.16227766e8  17.0  \\
            3.98107171e8  17.0  \\
            5.01187234e8  17.0  \\
            6.30957344e8  17.0  \\
            7.94328235e8  17.0  \\
            1.0e9  17.0  \\
            1.258925412e9  17.0  \\
            1.584893192e9  17.0  \\
            1.995262315e9  17.0  \\
            2.511886432e9  17.0  \\
            3.16227766e9  17.0  \\
            3.981071706e9  17.0  \\
            5.011872336e9  17.0  \\
            6.309573445e9  17.0  \\
            7.943282347e9  17.0  \\
            1.0e10  17.0  \\
        }
        ;
        \addlegendentry {limit $\nmalicious=5$}
\end{axis}
\end{tikzpicture}

%% file: Journal/sections/conclusion.tex
\section{Conclusion}
\label{sec:conclusion}
We analyzed the problem of distributed learning in the presence of Byzantine computation errors within a novel framework that extends the known gradient coding framework by an interactive communication between the \master and the workers and allows local computations at the \master. This framework has been analyzed with regard to its fundamental limits, and we derived the minimum number of local computations and a lower bound for the required communication from workers to the \master. By constructing and analyzing a particular scheme, we show how light interactive communication between the \master and the workers and verifying local computations at the \master can reduce required replication at the workers from $2\nmalicious+1$ to $\nmalicious+1$ in the presence of $\nmalicious$ malicious workers. It was shown that by a generalization of our previous scheme, it is possible to reduce the number of local computations even further. Namely, by considering a slightly higher replication of $\nmalicious+\nhonest$, the number of local computations decreases proportional to $1/\nhonest$. Alternatively, the higher replication can be exploited to additionally tolerate stragglers in a distributed computing setting. We showed that with a fractional repetition data assignment, the scheme achieves the optimal number of local computations at the \master. Future work includes the improvements of the converse and achievability bounds for the communication overhead, the generalization of the fundamental limits to a broader class of data assignments, and an investigation of the fundamental trade-off between uplink and downlink communication as well as their trade-off with the number of rounds.

%% file: Journal/sections/large_table.tex
\cref{tab:key_error_pattern_generalized} shows the symmetrization attack strategy for the general case where the replication is $\replfact = \nmalicious+\nhonest$ and $\nworker = \ngroup(\nmalicious+\nhonest)$. The table shows the case where all malicious workers are contained in the same fractional repetition group.
In order to avoid being detected by the \master through majority vote, the malicious workers split into subgroups of size $\nhonest$. In each subgroup, they generate their responses based on the same claimed partial gradient values. The claimed partial gradient values are chosen in the same way as explained for the case $\nhonest=1$ in \cref{sec:attack}.

\begin{table}[h]
  \caption{Claimed Partial Gradients for Symmetrization Attack.}
    \centering\resizebox{0.9\columnwidth}{!}{%
    \begin{tabularx}{1.1\columnwidth}{l|lllllll}
        & $\wpgrad{1}{\wind}$ & $\wpgrad{2}{\wind}$ 
        & $\dots$  & $\wpgrad{\compbound}{\wind}$ & $\wpgrad{\compbound+1}{\wind}$ & $\dots$ & $\wpgrad{\ngrad}{\wind}$ \\ 
        \addlinespace[3pt]
        \toprule
        $\worker[1]$ & $\pgrad[1]^{\prime\prime}$ & $\pgrad[2]^\prime$ & $\dots$  & $\pgrad[\lfloor \frac{\nmalicious}{\nhonest} \rfloor]^\prime$ & $\pgrad[\lfloor \frac{\nmalicious}{\nhonest} \rfloor + 1]^\prime$ & $\dots$ & $\pgrad[\ngrad]^\prime$ \\ 
        $\worker[2]$ & $\pgrad[1]^{\prime\prime}$ & $\pgrad[2]^\prime$ & $\dots$  & $\pgrad[\lfloor \frac{\nmalicious}{\nhonest} \rfloor]^\prime$ & $\pgrad[\lfloor \frac{\nmalicious}{\nhonest} \rfloor + 1]^\prime$ & $\dots$ & $\pgrad[\ngrad]^\prime$ \\ 
        $\vdots$ & $\vdots$ & $\vdots$ & $\vdots$ & $\vdots$ & $\vdots$ & $\dots$ & $\vdots$ \\
        $\worker[\nhonest]$ & $\pgrad[1]^{\prime\prime}$ & $\pgrad[2]^\prime$ & $\dots$  & $\pgrad[\lfloor \frac{\nmalicious}{\nhonest} \rfloor]^\prime$ & $\pgrad[\lfloor \frac{\nmalicious}{\nhonest} \rfloor + 1]^\prime$ & $\dots$ & $\pgrad[\ngrad]^\prime$ \\ 
        \addlinespace[3pt]
        \hdashline
        \addlinespace[3pt]
        $\worker[\nhonest+1]$ & $\pgrad[1]^\prime$ & $\pgrad[2]^{\prime\prime}$ & $\dots$  & $\pgrad[\lfloor \frac{\nmalicious}{\nhonest} \rfloor]^\prime$ & $\pgrad[\lfloor \frac{\nmalicious}{\nhonest} \rfloor + 1]^\prime$ & $\dots$ & $\pgrad[\ngrad]^\prime$ \\ 
        $\worker[\nhonest+2]$ & $\pgrad[1]^\prime$ & $\pgrad[2]^{\prime\prime}$ & $\dots$  & $\pgrad[\lfloor \frac{\nmalicious}{\nhonest} \rfloor]^\prime$ & $\pgrad[\lfloor \frac{\nmalicious}{\nhonest} \rfloor + 1]^\prime$ & $\dots$ & $\pgrad[\ngrad]^\prime$ \\ 
        $\vdots$ & $\vdots$ & $\vdots$ & $\vdots$ & $\vdots$ & $\vdots$ & $\dots$ & $\vdots$ \\
        $\worker[2\nhonest]$ & $\pgrad[1]^\prime$ & $\pgrad[2]^{\prime\prime}$ & $\dots$  & $\pgrad[\lfloor \frac{\nmalicious}{\nhonest} \rfloor]^\prime$ & $\pgrad[\lfloor \frac{\nmalicious}{\nhonest} \rfloor + 1]^\prime$ & $\dots$ & $\pgrad[\ngrad]^\prime$ \\ 
        \addlinespace[3pt]
        \hdashline
        \addlinespace[3pt]
        $\vdots$ & $\vdots$ & $\vdots$ & $\ddots$ & $\vdots$ & $\vdots$ & $\dots$ & $\vdots$ \\
        \addlinespace[3pt]
        \hdashline
        \addlinespace[3pt]
        $\worker[\lfloor \frac{\nmalicious}{\nhonest} \rfloor \nhonest - \nhonest +1]$ & $\pgrad[1]^\prime$ & $\pgrad[2]^\prime$ & $\dots$ & $\pgrad[\lfloor \frac{\nmalicious}{\nhonest} \rfloor]^{\prime\prime}$ & $\pgrad[\lfloor \frac{\nmalicious}{\nhonest} \rfloor + 1]^\prime$ & $\dots$ & $\pgrad[\ngrad]^\prime$ \\ 
        $\worker[\lfloor \frac{\nmalicious}{\nhonest} \rfloor \nhonest - \nhonest +2]$ & $\pgrad[1]^\prime$ & $\pgrad[2]^\prime$ & $\dots$ & $\pgrad[\lfloor \frac{\nmalicious}{\nhonest} \rfloor]^{\prime\prime}$ & $\pgrad[\lfloor \frac{\nmalicious}{\nhonest} \rfloor + 1]^\prime$ & $\dots$ & $\pgrad[\ngrad]^\prime$ \\ 
        $\vdots$ & $\vdots$ & $\vdots$ & $\vdots$ & $\vdots$ & $\vdots$ & $\dots$ & $\vdots$ \\
        $\worker[\lfloor \frac{\nmalicious}{\nhonest} \rfloor\nhonest]$ & $\pgrad[1]^\prime$ & $\pgrad[2]^\prime$ & $\dots$ & $\pgrad[\lfloor \frac{\nmalicious}{\nhonest} \rfloor]^{\prime\prime}$ & $\pgrad[\lfloor \frac{\nmalicious}{\nhonest} \rfloor + 1]^\prime$ & $\dots$ & $\pgrad[\ngrad]^\prime$ \\ 
        \addlinespace[3pt]
        \hdashline
        \addlinespace[3pt]
        $\worker[\lfloor \frac{\nmalicious}{\nhonest} \rfloor\nhonest +1]$ & $\pgrad[1]^\prime$ & $\pgrad[2]^\prime$ & $\dots$ & $\pgrad[\lfloor \frac{\nmalicious}{\nhonest} \rfloor]^{\prime}$ & $\pgrad[\lfloor \frac{\nmalicious}{\nhonest} \rfloor + 1]^\prime$ & $\dots$ & $\pgrad[\ngrad]^\prime$ \\ 
        $\worker[\lfloor \frac{\nmalicious}{\nhonest} \rfloor\nhonest +2]$ & $\pgrad[1]^\prime$ & $\pgrad[2]^\prime$ & $\dots$ & $\pgrad[\lfloor \frac{\nmalicious}{\nhonest} \rfloor]^{\prime}$ & $\pgrad[\lfloor \frac{\nmalicious}{\nhonest} \rfloor + 1]^\prime$ & $\dots$ & $\pgrad[\ngrad]^\prime$ \\ 
        $\vdots$ & $\vdots$ & $\vdots$ & $\vdots$ & $\vdots$ & $\vdots$ & $\dots$ & $\vdots$ \\
        $\worker[\nmalicious+\nhonest]$ & $\pgrad[1]^\prime$ & $\pgrad[2]^\prime$ & $\dots$  & $\pgrad[\lfloor \frac{\nmalicious}{\nhonest} \rfloor]^\prime$ & $\pgrad[\lfloor \frac{\nmalicious}{\nhonest} \rfloor + 1]^\prime$ & $\dots$ & $\pgrad[\ngrad]^\prime$ \\ 
    \end{tabularx}
  }
    \label{tab:key_error_pattern_generalized}
\end{table}

%% file: Journal/sections/algorithms.tex
\cref{alg:elimination_tournament} presents an algorithmic description of the elimination tournament described in \cref{sec:scheme}. It operates on the set $\groupset$ of disjoint groups of workers which agree on their initial responses. By running matches between group representatives successively, it finds the set $\elimset$ of malicious workers. As auxiliary functions, this algorithm requires the $match()$ function (\cref{alg:match}), a function $draw()$ that randomly selects one element from a set, a function $commit()$ which queries workers for the voting round, and a function $localComp()$ to compute a given partial gradient locally at the \master.

{
\ifarxiv \else \renewcommand{\baselinestretch}{1} \fi
\begin{algorithm}[h]
    \SetAlgoLined
    \SetKwInOut{Input}{Input}
    \SetKwInOut{Output}{Output}
    \SetKwInOut{Require}{Require}
    \Input{Workers $\worker[\indone]$ and $\worker[\indtwo]$, s.t. 
        $\noisyiresp_{0,\indone} \neq \noisyiresp_{0,\indtwo}$.
    }
    \Output{Value $\comp[\compind]{\wpgrad{\indcheck}{\indone}}$, index $\indcheck$, coordinate $\compind$.}
    $\gind_\mathrm{min} \gets 1$;
    $\gind_\mathrm{max} \gets \ngrad$;
    $\roundind \gets 0$\;
    $\compind \in \left\{ \compind^\prime \,\middle\vert\, \comp[\compind^\prime]{\noisyiresp_{0,\indone}} \neq \comp[\compind^\prime]{\noisyiresp_{0,\indtwo}} \right\}$\; 
    \While{$\gind_\mathrm{max} - \gind_\mathrm{min} > 0$}{
        $\roundind \gets \roundind + 1$;
        $\gind_\mathrm{half} \gets \gind_\mathrm{min} + \lceil \frac{\gind_\mathrm{max} - \gind_\mathrm{min}}{2} \rceil$\;
        $\textbf{request } \iresp_{\roundind,\indone} \gets
            \sum_{\gind=\gind_\mathrm{min}}^{\gind_{\mathrm{half}}}
            \comp{ \wpgrad{\gind}{\indone} }$ from $\worker[\indone]$\;
        $\voteset \gets commit\big( \noisyiresp_{\roundind,\indone},\worker[\indtwo],\compind,\gind_\mathrm{min}:\gind_\mathrm{max} \big)$\;
        
        \eIf{$\card{\voteset} > 0$}{
            $\gind_\mathrm{min} \gets \gind_\mathrm{half}$\;
        }{
            $\gind_\mathrm{max} \gets \gind_\mathrm{half}$\;
        }
    }
    $\indcheck \gets \gind_\mathrm{min}$;
    $\comp[\compind]{\wpgrad{\indcheck}{\indone}} \gets \iresp_{\roundind,\indone}$\;
    \caption{Match between two workers.}
    \label{alg:match}
\end{algorithm}}
\clearpage

{
\ifarxiv \else \renewcommand{\baselinestretch}{1} \fi
\begin{algorithm}[h]
    \SetAlgoLined
    \SetKwInOut{Input}{Input}
    \SetKwInOut{Output}{Output}
    \SetKwInOut{Require}{Require}
    \Input{Set $\groupset$ of disjoint worker groups, where $\workerset \in \groupset$ has $\workerset \subset \range{\nworker}$ and $\nhonest \leq \card{\workerset} \leq \nmalicious$.}
    \Require{$match(), draw()$, $commit()$, $localComp()$.}
    \Output{Set $\elimset$ of malicious workers.}
    
    $\elimset \gets \left\{ 1,2,\dots,\nworker \right\} \setminus \cup_{\workerset \in \groupset} \workerset$\;
    \While{$\card{\groupset} > 1$}{
        $\workerset_1 \gets draw\left(\groupset\right)$;
        $\worker[1] \gets draw\left(\workerset_1\right)$\;
        $\workerset_2 \gets draw\left(\groupset \setminus \left\{ \workerset_1 \right\}\right)$;
        $\worker[2] \gets draw\left(\workerset_2\right)$\;
        $\comp{\wpgrad{\gind}{1}},\gind,\compind \gets match(\worker[1],\worker[2])$ (\texttt{\cref{alg:match}})\;
        $\voteset_1 \gets \{ \worker[1] \}$\;
        $\voteset_2 \gets commit\big( \comp{\wpgrad{\gind}{1}},\compind,\gind,\workerset_1 \cup \workerset_2 \setminus \{ \worker[1],\worker[2] \} \big)$\;
        $\voteset \gets \voteset_1 \cup \voteset_2$\;
        \uIf{$\card{\voteset} < \nhonest$}{
            $\elimset \gets \elimset \cup \voteset$\;
            $\workerset_1 \gets \workerset_1 \setminus \voteset$;
            $\workerset_2 \gets \workerset_2 \setminus \voteset$\;
        }
        \uElseIf{$\card{\voteset} > \card{\workerset_1} + \card{\workerset_2} - \nhonest$}{
            $\elimset \gets \elimset \cup \workerset_1 \cup \workerset_2 \setminus \voteset$\;
            $\workerset_1 \gets \workerset_1 \cap \voteset$;
            $\workerset_2 \gets \workerset_2 \cap \voteset$\;
        }
        \Else{
            $\comp{\pgrad[\gind]} \gets localComp(\gind,\compind)$\;
            \eIf{$\comp{\wpgrad{\gind}{1}} \neq \comp{\pgrad[\gind]}$}{
                $\elimset \gets \elimset \cup \voteset$\;
                $\workerset_1 \gets \workerset_1 \setminus \voteset$;
                $\workerset_2 \gets \workerset_2 \setminus \voteset$\;
            }{
                $\elimset \gets \elimset \cup \workerset_1 \cup \workerset_2 \setminus \voteset$\;
                $\workerset_1 \gets \workerset_1 \cap \voteset$;
                $\workerset_2 \gets \workerset_2 \cap \voteset$\;
            }
        }
        \For{$i=1,2$}{
            \If{$\card{\workerset_i} < \nhonest$}{
                $\elimset \gets \elimset \cup \workerset_i$\;
                $\groupset \gets \groupset \setminus \left\{ \workerset_i \right\}$\;
            }
        }
    }
    \caption{Elimination Tournament.}
    \label{alg:elimination_tournament}
\end{algorithm}}